\documentclass[preprint,12pt]{elsarticle}
\usepackage{graphicx}

\usepackage{amssymb}
 \usepackage{amsthm}
\usepackage{amsmath}
\usepackage{lineno}

\usepackage{enumerate}
\usepackage{here}
\usepackage{indentfirst}
\usepackage{bm}
\usepackage{url}
\usepackage{color}
\usepackage{soul}
\usepackage{here}
\usepackage{tikz}
\usetikzlibrary{shapes.geometric}
\usetikzlibrary {shapes.misc}
\usetikzlibrary{positioning}
\usepackage[hang,small,bf]{caption}
\usepackage[subrefformat=parens]{subcaption}
\def\XXint#1#2#3{{\setbox0=\hbox{$#1{#2#3}{\int}$ }
		\vcenter{\hbox{$#2#3$ }}\kern-.6\wd0}}

\newcommand{\pd}{p'}
\newcommand{\pdm}{{p'}^{(m)}}
\newcommand{\pda}{{p'}^{(a)}}
\newcommand{\pdb}{{p'}^{(b)}}
\newcommand{\hatp}{\hat{p'}}
\newcommand{\hatpm}{\hat{p'}^{(m)}}
\newcommand{\hatpa}{\hat{p'}^{(a)}}
\newcommand{\hatpb}{\hat{p'}^{(b)}}
\newcommand{\qd}{q'}
\newcommand{\qdm}{{q'}^{(m)}}
\newcommand{\qda}{{q'}^{(a)}}
\newcommand{\qdb}{{q'}^{(b)}}
\newcommand{\hatq}{\hat{q'}}
\newcommand{\hatqm}{\hat{q'}^{(m)}}
\newcommand{\hatqa}{\hat{q'}^{(a)}}
\newcommand{\hatqb}{\hat{q'}^{(b)}}
\newcommand{\uvec}{\bm{u}}
\newcommand{\uvecm}{\bm{u}^{(m)}}
\newcommand{\uveca}{\bm{u}^{(a)}}
\newcommand{\uvecb}{\bm{u}^{(b)}}
\newcommand{\hatu}{\hat{\bm{u}}}
\newcommand{\hatum}{\hat{\bm{u}}^{(m)}}
\newcommand{\hatua}{\hat{\bm{u}}^{(a)}}
\newcommand{\hatub}{\hat{\bm{u}}^{(b)}}
\newcommand{\vvec}{\bm{v}}
\newcommand{\vvecm}{\bm{v}^{(m)}}
\newcommand{\vveca}{\bm{v}^{(a)}}
\newcommand{\vvecb}{\bm{v}^{(b)}}
\newcommand{\hatv}{\hat{\bm{v}}}
\newcommand{\hatvm}{\hat{\bm{v}}^{(m)}}
\newcommand{\hatva}{\hat{\bm{v}}^{(a)}}
\newcommand{\hatvb}{\hat{\bm{v}}^{(b)}}
\newcommand{\rhom}{\rho^{(m)}}
\newcommand{\rhoa}{\rho^{(a)}}
\newcommand{\rhob}{\rho^{(b)}}
\newcommand{\Km}{K^{(m)}}

\newcommand{\nvec}{\bm{n}}
\newcommand{\nvecm}{\bm{n}^{(m)}}
\newcommand{\nveca}{\bm{n}^{(a)}}
\newcommand{\nvecb}{\bm{n}^{(b)}}
\newcommand{\sigmam}{\sigma^{(m)}}
\newcommand{\sigmaa}{\sigma^{(a)}}
\newcommand{\sigmab}{\sigma^{(b)}}

\newcommand{\jump}[1]{\ensuremath{[\![#1]\!]} }

\journal{Elsevier}

\begin{document}

\begin{frontmatter}



\title{
Level set-based shape optimization of deformable structures for manipulating sound propagation
}


\author[label1,label2]{Yuki~Noguchi\corref{cor1}} 
\ead{noguchi@mech.t.u-tokyo.ac.jp}
\cortext[cor1]{Corresponding author.
	Tel.: +81-3-5841-0294;
	Fax: +81-3-5841-0294.}
\author[label1,label2]{Takayuki~Yamada}

\address[label1]{Department of Strategic Studies, Institute of Engineering Innovation, Graduate School of Engineering, The University of Tokyo, Yayoi 2-11-16, Bunkyo--ku, Tokyo 113-8656, Japan.}
\address[label2]{Department of Mechanical Engineering, Graduate School of Engineering, The University of Tokyo, Yayoi 2-11-16, Bunkyo--ku, Tokyo 113-8656, Japan.}

\begin{abstract}
In this paper, we propose a level set-based shape optimization method for acoustic wave propagation problems with a deformable structure. 
First, we propose a mathematical model for acoustic wave propagation with a deformed structure based on coordinate transformation and the Eulerian approach. 
Next, we formulate the shape optimization problem and perform sensitivity analysis based on the shape derivative concept. 
We then construct an optimization algorithm based on the framework of a level set-based shape and topology optimization method. 
Finally, we provide two-dimensional optimization examples that demonstrate the effectiveness of our proposed method in providing optimized designs with desired functionality, while considering the structural deformation. 
\end{abstract}


\end{frontmatter}


\section{Introduction}\label{intro}
Structural optimization, which encompasses size, shape, and topology optimization, has been widely applied in various fields, particularly in structural mechanics. 
Among these methods, shape and topology optimization have earned considerable attention from researchers and engineers due to their ability to generate flexible designs that maximize structural performance. 
With the recent emergence of additive manufacturing, the optimized structures obtained by these methods have become increasingly feasible \cite{liu2018current}. The procedure of additive manufacturing can be incorporated into the optimization method \cite{dapogny2019shape}, thereby expanding their practical applications. 

Since the pioneering work of Bends{\o}e and Kikuchi \cite{bendsoe1988generating}, various established approaches have been proposed, including the density-based approach \cite{bendsoe1989optimal,bendsoe2004topology} and the level set-based methods \cite{allaire2004structural,yamada2010topology}. 
Initially, these methods were applied to linear elasticity problems, but researchers have since expanded their scope to more complex problems, such as nonlinear elasticity \cite{jung2004topology} and fluid dynamics \cite{borrvall2003topology}. 
Acoustic wave propagation has also been a target of structural optimization. 
For example, B\"{a}ngtsson et al. \cite{bangtsson2003shape} proposed a shape optimization method for designing an acoustic horn that efficiently transmits sounds. Lately, Wadbro and Berggren \cite{wadbro2006topology} conducted topology optimization for the optimal design of horns. 
D{\"u}hring et al. \cite{duhring2008acoustic} conducted topology optimization to obtain structures to reduce sound noises. 
Although these works focused on airborne sounds, the interaction between airborne sounds
and structural oscillations has also been incorporated into the structural optimization method, as shown in a study by Shu et al. \cite{shu2014level}. 
The effect of viscous and thermal boundary layers has also been included in shape \cite{andersen2019shape,tissot2020optimal} and topology optimization  \cite{christensen2017topology,noguchi2021topologySLNS}. 
Recently, acoustic metamaterials \cite{liu2000locally} have earned researchers' interest due to their unusual acoustic properties, such as the negative refractive index \cite{li2004double}, which enables them to bend and focus sound waves in ways that are impossible with conventional materials.
Structural optimization methods have been applied for the optimal design of acoustic metamaterials \cite{lu2013topology,yang2016effective,dong2019systematic}.

All of the aforementioned examples are aimed at maximizing acoustic performance in a passive manner, assuming that a structural configuration is fixed. However, structures are often subjected to deformation due to external factors such as mechanical loads, heat sources, and electromagnetic fields. The acoustic properties of a deformed structure are different from those of an undeformed one, which means that an optimized design may not exhibit the desired acoustic performance when subjected to deformation. 
Therefore, structural deformation caused by external factors must be included in the design to achieve the robust functionality of the structure. 
In contrast, changes in acoustic performance when the structure undergoes deformation could realize a multifunctional acoustic device that shows different functionalities depending on its deformation. 
A prominent example is programmable metamaterials \cite{cui2014coding,yang2017programmable}, which possess tailored properties against external forces via the deformation of the unit cell. 
Therefore, an optimal design method that includes the structural deformation enables us to realize multiple desired acoustic performances, and more fruitful functionalities of the optimized design can be expected compared to the fixed geometry case. 

Related works of optimal design focusing on the deformation of the structure can be found in the applications of soft robotics and 4-D printing. There have been topology optimization methods targeting structures that deform due to actuators or heat sources \cite{geiss2019combined,sato2023topology}. 
However, they aimed to realize a certain deformation or movement of the structures, and they did not target responses in other physical fields at the deformed geometry.
The research examples of structural optimization methods for wave propagation problems under consideration of structural deformation have been limited for now. 
Dalklint et al. \cite{dalklint2022tunable} proposed a topology optimization method for a unit cell of a phononic crystal under uniaxial and biaxial extension to obtain a tunable bandgap structure. 
To achieve the tunable bandgap metamaterial design, Luo and Li \cite{luo2022tunable} proposed a gradient-free topology optimization algorithm based on the material-field series-expansion method.
However, these methods are based on the assumption that the phononic crystal is made of unit cells that are infinitely periodically arrayed, and that each unit cell deforms uniformly in a macroscopic view.

In this research, we propose a shape optimization method for acoustic wave propagation problems with a deformable structure. Unlike \cite{dalklint2022tunable,luo2022tunable}, which target periodic structures, we consider structural deformation expressed with a non-uniform displacement field that satisfies certain boundary conditions. 
First, we propose a mathematical model for acoustic wave propagation with a deformed structure based on coordinate transformation. Based on this model, acoustic performance with the deformed geometry can be numerically analyzed using the finite element method (FEM) with the Eulerian approach. 
The Eulerian approach is useful because there is no need to deform the mesh to analyze the wave propagation with the deformed geometry. In addition, sensitivity analysis can be conducted as in the case without deformation. 
Next, we formulate the shape optimization problem within the framework of the level set-based method.
In particular, we introduce a level set method proposed by Feppon et al. \cite{feppon2017introducing}. In their method, a reaction--diffusion equation is used to update the shape represented by a level set function to control the perimeter. 
We introduce this method because the use of a reaction--diffusion equation rather than the Hamilton--Jacobi equation in the classic level set-based shape and topology optimization \cite{allaire2004structural}, makes numerical treatment in the FEM easier. 

Similar approaches that use coordinate transformation for numerical analysis with the Eulerian mesh can be found in fluid-solid interaction problems \cite{dunne2006eulerian, WICK201314}.
It has also been reported that the shape derivative, which is the design index for shape optimization, can be derived based on the model \cite{calisti2022shape}. 
However, there are no reports on acoustic wave propagation problems, and structural optimization has not been conducted for this type of problem. 

The remainder of this paper is organized as follows. 
In Section~\ref{sec: Design settings}, we state an acoustic wave propagation problem with a deformable structure and formulate weak forms that are used in the FEM analyses. 
Based on these formulations, the shape optimization problem is formulated in a general form
in Section~\ref{sec: Formulation of Topology optimization}. 
Section~\ref{sec: Sensitivity analysis} provides the details of sensitivity analysis for the shape derivative. 
In Section~\ref{sec: Level set-based topology optimization}, we introduce a level set-based shape and topology optimization proposed by Feppon et al. \cite{feppon2017introducing} to solve the optimization problem. 
Several notes on the numerical implementation are listed in Section~\ref{sec:Numerical implementation} including the optimization algorithm. 
We provide two-dimensional numerical examples in Section~\ref{sec: numerical examples} to confirm the effectiveness of the proposed optimization method. 
Two kinds of problems to maximize the transmission loss of a tube and to design an acoustic lens are provided, and we show how the proposed optimal design method works with structural deformation. 
Finally, we conclude this work in Section~\ref{sec: Conclusion}.

\section{Acoustic wave propagation with a deformable structure}\label{sec: Design settings}
\subsection{Mathematical model for acoustic pressure}\label{sec: Mathematical model for acoustic pressure}
In this section, we briefly introduce a mathematical model for an acoustic wave propagation problem.
\begin{figure}[h!]
	\begin{center}
		\includegraphics[scale=0.5]{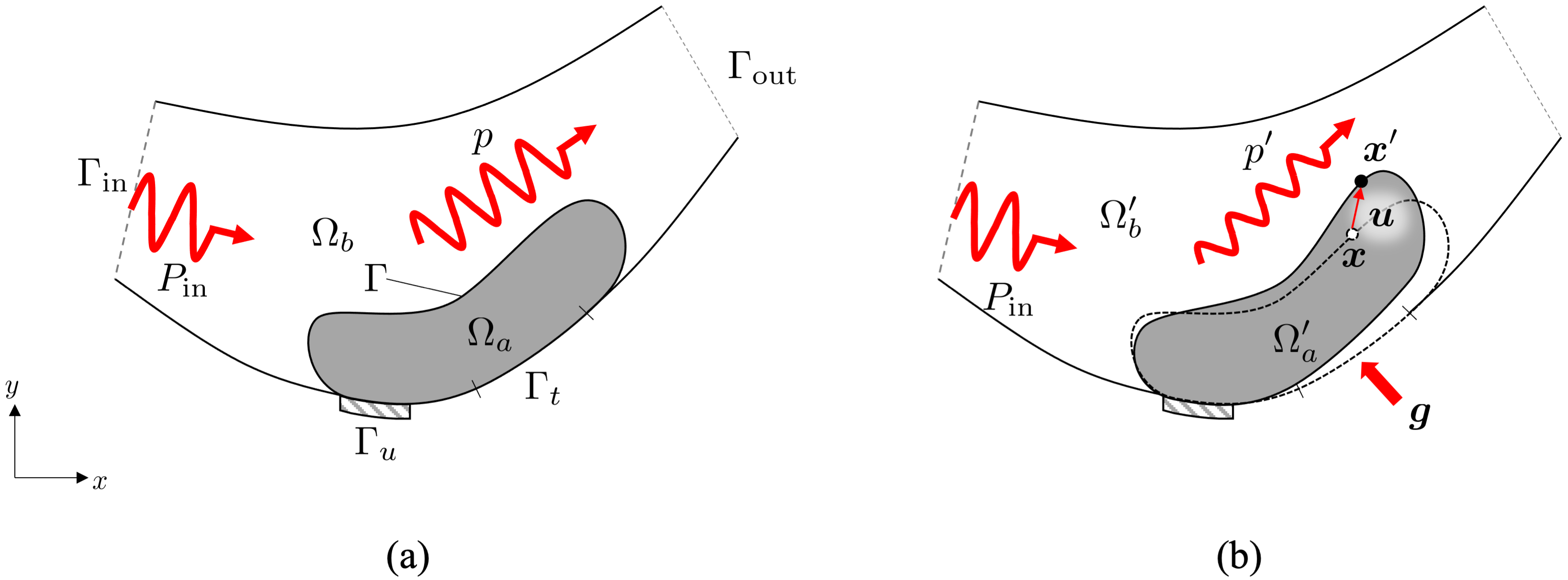}
	\end{center}
	\caption{Geometry and boundary conditions. 
	(a) Acoustic wave propagation without deformation. 
	(b) Acoustic wave propagation with deformation. 
}
	\label{fig: GeomConcept}
\end{figure}

The geometry and boundary conditions of the acoustic wave propagation problem are depicted in Fig.~\ref{fig: GeomConcept}.
Specifically, as illustrated in Fig.~\ref{fig: GeomConcept}(a), 
we aim to study wave propagation in a two-phase material system, comprised of two distinct domains, $\Omega_a$ and $\Omega_b$.
The domain $\Omega_a$ represents the solid material, while $\Omega_b$ corresponds to the air-filled region. The interface between the two domains is represented by $\Gamma$. 
An inlet boundary condition, $\Gamma_{\mathrm{in}}$, is imposed on a portion of $\partial \Omega_b$, where an incident wave is directed towards the system. 
The incident wave subsequently propagates through the two domains, $\Omega_a$ and $\Omega_b$, and ultimately reaches the outlet boundary, $\Gamma_{\mathrm{out}}$, which is connected to free space.

We consider time-harmonic wave propagation with angular frequency $\omega$. Additionally, we only consider longitudinal wave motion in both air and elastic medium, which is justifiable when the impedance ratio between them is high. The boundary value problem for the acoustic wave propagation problem is then formulated as follows:
\begin{align}
    -\nabla \cdot \left(\frac{1}{\rho(\bm{x})} \nabla p \right) - \frac{\omega^2}{K(\bm{x})}p &= 0~~~\mathrm{in}~\Omega_{\mathrm{all}},
    \label{eq: Helmholtz in strong}\\
    %
    %
    \bm{n}\cdot \left(\frac{1}{\rho_b} \nabla p \right) + \frac{i k_b}{\rho_b}p &=  \left(1- \frac{\bm{k}\cdot \nvec}{k_b}\right)\frac{ik_b}{\rho_b}P_{\mathrm{in}}~~~\mathrm{on}~\Gamma_{\mathrm{in}},
    \label{eq: Inlet}\\
    \bm{n}\cdot \left(\frac{1}{\rho_b} \nabla p \right) + \frac{i k_b}{\rho_b}p &= 0~~~\mathrm{on}~\Gamma_{\mathrm{out}},
    \label{eq: Outlet}\\
    \bm{n}\cdot \left(\frac{1}{\rho(\bm{x})} \nabla p \right) &= 0~~~\mathrm{on}~\partial \Omega_{\mathrm{all}}\setminus (\Gamma_{\mathrm{in}}\cup \Gamma_{\mathrm{out}}),
    \label{eq: Sound-hard}
\end{align}
where $p$ denotes the complex amplitude of the acoustic pressure. $\Omega_{\mathrm{all}}=\Omega_a\cup \Omega_b$ represents the entire domain, and $\nvec$ is the outward unit normal vector. 
The mass density and bulk modulus are represented by 
$\rho(\bm{x})$ and $K(\bm{x})$, respectively, and they show a piecewise distribution depending on the position $\bm{x}$, as follows:
\begin{eqnarray}
	\rho(\bm{x})=
	\left\{
	\begin{array}{ll}
		\rho_a &\mathrm{if}~~\bm{x} \in \Omega_a\\
		\rho_b &\mathrm{if}~~\bm{x} \in \Omega_b
	\end{array}
	\right. ,
\end{eqnarray}
\begin{eqnarray}
	K(\bm{x})=
	\left\{
	\begin{array}{ll}
		K_a &\mathrm{if}~~\bm{x} \in \Omega_a\\
		K_b &\mathrm{if}~~\bm{x} \in \Omega_b
	\end{array}
	\right. ,
\end{eqnarray}
where $\rho_a$ and $\rho_b$ represent the mass densities of solid material and the air, respectively. 
Similarly, $K_a$ and $K_b$ are the bulk moduli of solid material and the air, respectively. 
Eq.~(\ref{eq: Inlet}) represents an incident wave condition on $\Gamma_{\mathrm{out}}$. 
$k_b$ is the wavenumber in the air, and $P_{\mathrm{in}}$ is the incident wave. 
We consider the input of a plane wave $P_{\mathrm{in}} = \exp{(-i\bm{k}\cdot \bm{x})}$ with a given wavevector $\bm{k}$. 
Eq.~(\ref{eq: Outlet}) represents an outlet condition on $\Gamma_{\mathrm{out}}$. A non-reflecting boundary condition is applied to this boundary. 
A sound-hard boundary condition is applied to the other external boundary of $\Omega_{\mathrm{all}}$ than the inlet and outlet boundaries, as shown in Eq.~(\ref{eq: Sound-hard}). 

Based on Eq.~(\ref{eq: Helmholtz in strong})--Eq.~(\ref{eq: Sound-hard}), the following weak form for $p$ can be formulated:
\begin{align}
	&\int_{\Omega_\mathrm{all}} \left(\frac{1}{\rho(\bm{x})}\nabla p \cdot \nabla \tilde{p} - \frac{\omega^2}{K(\bm{x})}p\tilde{p}\right)
	d\Omega
	+ \int_{\Gamma_\mathrm{in}\cup \Gamma_\mathrm{out}}\frac{i k_b}{\rho_b}  p \tilde{p}  d\Gamma\nonumber\\
	&=\int_{\Gamma_\mathrm{in}} \left(1-\frac{\bm{k}\cdot \bm{n}}{k_b}\right)\frac{i k_b}{\rho_a}P_\mathrm{in}\tilde{p}d\Gamma
	~~~\forall \tilde{p} \in H^1(\Omega_\mathrm{all}),
	\label{eq: Weak form of p}
\end{align}
where $\tilde{p}$ is a test function defined in the Sobolev space $H^1(\Omega_\mathrm{all})$.

We now consider the deformation of the structural domain $\Omega_a$ and its effect on acoustic wave propagation.
Due to the deformation, each point $\bm{x}$ in $\Omega_a$ is displaced to a new material point $\bm{x}'=\bm{x} + \uvec(\bm{x})$ by a given displacement vector $\uvec$, as shown in Fig.~\ref{fig: GeomConcept}(b). 
The resulting deformed structural domain $\Omega_a'$ can be expressed as $\Omega_a'=(\mathrm{Id}+\uvec)(\Omega_a)$ using the identity mapping $\mathrm{Id}$. 
Similarly, the interface $\Gamma$ deforms as $(\mathrm{Id}+\uvec)(\Gamma)$. 

The air-filled domain $\Omega_b'$ also deforms as a result of the structural deformation. 
We define the deformed air-filled domain $\Omega_b'$ as $\Omega_b' = (\mathrm{Id}+\uvec)(\Omega_b)$ by extending the displacement field $\uvec$ in $\Omega_a$ to the entire domain $\Omega_{\mathrm{all}}$.
We assume that $\uvec$ is sufficiently smooth, allowing for the inverse mapping to be defined. 
Additionally, we assume that $\uvec=0$ on the boundaries $\partial \Omega_b\setminus \Gamma$ 
to ensure that the external boundaries of the acoustic region remain undeformed.

Let $p'$ denote the acoustic pressure with the deformed configuration in which the same boundary conditions as Eq.~(\ref{eq: Inlet}), Eq.~(\ref{eq: Outlet}), and Eq.~(\ref{eq: Sound-hard}) are applied. 
The weak form satisfied by $\pd$ and defined by the material coordinate $\bm{x'}$ is given by: 
\begin{align}
	&\int_{\Omega_\mathrm{all}' } \left(\frac{1}{\rho(\bm{x'})}\nabla' \pd \cdot \nabla' \tilde{p} - \frac{\omega^2}{K(\bm{x'})}\pd \tilde{p}\right)
	d\Omega(\bm{x'})
	+ \int_{\Gamma_\mathrm{in}'\cup \Gamma_\mathrm{out}'}\frac{i k_b}{\rho_b}  \pd \tilde{p}  d\Gamma(\bm{x'})\nonumber\\
	&=\int_{\Gamma_\mathrm{in}'} \left(1-\frac{\bm{k}\cdot \bm{n}}{k_b}\right)\frac{i k_b}{\rho_b}P_\mathrm{in}\tilde{p}d\Gamma(\bm{x'}) ~~~\forall \tilde{p} \in H^1(\Omega_\mathrm{all}'),
	\label{eq: Weak form of p' before transformation}
\end{align}
where the symbol $\nabla'$ represents the spatial gradient with respect to $\bm{x'}$. 
The deformed domains are contained in the entire region $\Omega_\mathrm{all}' = \Omega_a' \cup \Omega_e'$.
Similarly, the deformed inlet and outlet boundaries are denoted by 
$\Gamma_\mathrm{in}'$ and $\Gamma_\mathrm{out}'$, respectively. 
However, these boundaries remain identical to their original positions due to the assumption that the displacement field $\uvec$ is equal to zero on $\Gamma_{\mathrm{in}}\cup\Gamma_{\mathrm{out}}$.

Eq.~(\ref{eq: Weak form of p' before transformation}) can be reformulated with respect to the spatial coordinate $\bm{x}$ by introducing the deformation gradient denoted as $\bm{F}(\bm{u}) = \frac{\partial \bm{x'}}{\partial \bm{x}} = \bm{I} + \nabla \bm{u}$, where $\bm{I}$ represents the second-order identity tensor. 
Subsequently, Eq. (\ref{eq: Weak form of p' before transformation}) can be transformed to the following form: 
\begin{align}
	&\int_{\Omega_\mathrm{all}} \left\{\frac{1}{\rho(\bm{x})} (\bm{F}^{-T}(\bm{u})\nabla \pd) \cdot (\bm{F}^{-T}(\bm{u})\nabla \tilde{p}) - \frac{\omega^2}{K(\bm{x})}p'\tilde{p}\right\}
	\det \bm{F(\bm{u})}d\Omega\nonumber\\
	&+ \int_{\Gamma_\mathrm{in}\cup \Gamma_\mathrm{out}}\frac{i k_b}{\rho_b}  \pd \tilde{p}  d\Gamma
  =\int_{\Gamma_\mathrm{in}} \left(1-\frac{\bm{k}\cdot \bm{n}}{k_b}\right)\frac{i k_b}{\rho_b}P_\mathrm{in}\tilde{p}d\Gamma,
	\label{eq: Weak form of p' after transformation}
\end{align}
where the boundary terms in Eq.~(\ref{eq: Weak form of p' after transformation}) are in the same form as those in Eq.~(\ref{eq: Weak form of p}) owing to the assumption that $\bm{u}=0$ on both $\Gamma_\mathrm{in}$ and  $\Gamma_\mathrm{out}$. 
Additionally, it is noteworthy that the solution $p'$ of Eq.~(\ref{eq: Weak form of p' after transformation}) is identical to $p$ in Eq.~(\ref{eq: Weak form of p}) when there is no deformation ($\uvec=0$) throughout the entire region $\Omega_{\mathrm{all}}$ since $\bm{F}(\uvec)=\bm{I}$ holds.

\subsection{Boundary value problem for the displacement field}\label{sec: Boundary value problem for the displacement field}
In Section \ref{sec: Mathematical model for acoustic pressure}, we treated the displacement field $\uvec$ as a given field. 
In this section, we define the boundary value problem for $\uvec$ that corresponds to the physical situation of the deformation. 

As a typical example, we consider a deformation under a traction force applied on a boundary of the structural domain $\Omega_a$, as shown in Fig.~\ref{fig: GeomConcept}(b).
We assume that the solid material in $\Omega_a$ is an isotropic and linear elastic material. 
A traction force is applied on $\Gamma_t$, while the displacement is fixed on $\Gamma_u$. 
The traction-free condition is applied on $\partial \Omega_a \setminus (\Gamma\cup \Gamma_t \cup \Gamma_u)$.
Then, we can formulate the boundary value problem for $\uvec$ as follows: 
\begin{align}
    -\nabla \cdot \sigma^{(a)}(\uvec) &= 0~~~\mathrm{in~}\Omega_a, \label{eq: Elasticity_orig}\\
    \sigma^{(a)}(\uvec)\cdot \nvec &= \bm{g}~~~\mathrm{on~}\Gamma_t,\label{eq: Traction}\\
    \sigma^{(a)}(\uvec)\cdot \nvec &= 0~~~\mathrm{on~}\partial \Omega_a \setminus (\Gamma\cup \Gamma_t \cup \Gamma_u),\label{eq: Traction-free}\\
    \uvec &= 0~~~\mathrm{on~}\Gamma_u.\label{eq: fixed}
\end{align}
Here, $\sigma^{(a)}(\bm{u}) = \bm{C}^{(a)}:\varepsilon(\bm{u})$ represents the stress tensor, where $\varepsilon(\bm{u})=(\nabla \bm{u} + (\nabla \bm{u})^T )/2$ is the strain tensor.
The elastic tensor of the solid material is denoted by $\bm{C}^{(a)}$. 
Eq.~(\ref{eq: Traction}) imposes the traction boundary condition with a given traction vector $\bm{g}$.
Eq.~(\ref{eq: Traction-free}) presents the traction-free condition on $\partial \Omega_a \setminus (\Gamma\cup \Gamma_t \cup \Gamma_u)$.
Eq.~(\ref{eq: fixed}) indicates the fixed constraint on $\Gamma_u$. 
The weak form corresponding to Eq.~(\ref{eq: Elasticity_orig})--Eq.~(\ref{eq: fixed}) is formulated as follows: 
\begin{align}
	\int_{\Omega_a} \sigma^{(a)}(\bm{u}) : \varepsilon (\bm{\tilde{u}}) d\Omega = \int_{\Gamma_t}\bm{g}\cdot \bm{\tilde{u}}d\Gamma~~~\forall \bm{\tilde{u}}\in V(\Omega_a), 
	\label{eq: elasticity_original}
\end{align}
where $V(\Omega_a) = \left\{ \bm{\tilde{u}}    \in \{H^1(\Omega_a)\}^d  ~|~   \bm{\tilde{u}} = 0~\mathrm{on~}\Gamma_u     \right\}$ denotes the functional space of a test function $\bm{\tilde{u}}$ with the number of spatial dimensions $d$.

As described in Section~\ref{sec: Mathematical model for acoustic pressure}, 
to formulate Eq.~(\ref{eq: Weak form of p' after transformation}), it is necessary to define the displacement field $\uvec$ not only in $\Omega_a$ but also in $\Omega_b$. 
To accomplish this, we introduce the Ersatz approach, 
which models a void region as an elastic medium with a small Young's modulus. 
It should be noted that defining $\uvec$ in $\Omega_b$ is not limited to this method; 
one alternative is to extrapolate the solution of Eq.~(\ref{eq: elasticity_original}), which has been employed to solve fluid-solid interaction problems in the fully Eulerian framework \cite{YEO2020109482}. We have chosen the Ersatz approach for its simplicity. 

In our formulation, we treat the air-filled region $\Omega_b$ as a void region and model it using a weak elastic material. 
Furthermore, fixed constraints are imposed not only on $\Gamma_u$ but also on $\partial \Omega_b \setminus \Gamma$ to ensure that $\uvec=0$ on these boundaries. 
With these considerations, the weak form in Eq.~(\ref{eq: elasticity_original}) can be modified as follows:
\begin{align}
 \int_{\Omega_\mathrm{all}} \sigma(\uvec) : \varepsilon (\bm{\tilde{u}}) d\Omega = 
 \int_{\Gamma_t}\bm{g}\cdot \bm{\tilde{u}}d\Gamma~~~\forall \bm{\tilde{u}}\in W(\Omega_\mathrm{all}), 
	\label{eq: elasticity_Ersatz}
\end{align}
where $W(\Omega_\mathrm{all}) = \left\{ \bm{\tilde{u}}    \in \{H^1(\Omega_\mathrm{all})\}^d  ~|~   \bm{\tilde{u}} = 0~\mathrm{on~} \Gamma_u \cup  (\partial\Omega_b\setminus\Gamma) \right\}$. 
The stress tensor $\sigma(\uvec)$ is defined as $\bm{C}(\bm{x}):\varepsilon(\uvec)$, where $\bm{C}(\bm{x})$ is the elasticity tensor with a piecewise constant distribution, given by:
\begin{eqnarray}
	\bm{C}(\bm{x})=
	\left\{
	\begin{array}{ll}
		\bm{C}^{(a)} &\mathrm{if}~~\bm{x} \in \Omega_a \\
  		\bm{C}^{(b)} = \varepsilon_u\bm{C}^{(a)} &\mathrm{if}~~\bm{x} \in \Omega_b
	\end{array}
	\right. ,\label{eq: Elasticity tensor}
\end{eqnarray}
where $\varepsilon_u$ is a small positive coefficient used to express the weak material in $\Omega_b$, and $\varepsilon_u \ll 1$.

The procedure for obtaining $\pd$ consists of two steps based on these formulations. 
First, $\uvec$ is obtained by solving Eq.~(\ref{eq: elasticity_Ersatz}) in $\Omega_{\mathrm{all}}$. Next, Eq.~(\ref{eq: Weak form of p' after transformation}) is solved using the obtained $\uvec$.

\section{Formulation of the optimization problem}\label{sec: Formulation of Topology optimization}

In this section, we present the general framework of the optimization problem for optimizing  the shape of the structural domain $\Omega_a$ to minimize an objective function $J$, which measures the performance of the structure in controlling the behaviors of acoustic waves. 
We assume that the objective function $J$ depends implicitly on $\Omega_a$ through two state variables, namely the acoustic pressure $\pd$ and the displacement vector $\uvec$.
To emphasize this dependence, we use the notation $\pd(\Omega_a)$ and $\uvec(\Omega_a)$.
Then, the optimization problem can be formulated as follows: 
\begin{equation}
	\begin{aligned}
		&\min_{\Omega_a}
		&& J\left( \pd(\Omega_a),\uvec(\Omega_a)\right)&& \\
		&{\rm subject~to}
		&&{\rm Governing~equation~for~}\pd,\\
		& &&{\rm Governing~equation~for~}\bm{u}.\\
	\end{aligned}
	\label{eq: topopt}
\end{equation}
The detailed definition of the objective function $J$ is provided in Section~\ref{sec: numerical examples}. 
Note that, even if $J$ depends only on $\pd(\Omega_a)$, the optimization problem is still constrained by both governing equations for $\pd$ and $\uvec$ due to the dependence of $\pd$ on $\uvec$ through Eq.~(\ref{eq: Weak form of p' after transformation}).

\section{Sensitivity analysis}\label{sec: Sensitivity analysis}
In order to solve the optimization problem presented in Eq.~(\ref{eq: topopt}), 
it is necessary to calculate design sensitivity. 
To optimally design the shape of the structural domain $\Omega_a$, we introduce the concept of shape derivative.
The shape derivative represents the variation of the objective function when the structural domain undergoes a slight deformation. 
In our case, the shape derivative measures the variation of $J$ when the interface $\Gamma$ between $\Omega_a$ and $\Omega_b$ moves slightly, while the entire domain $\Omega_{\mathrm{all}}$ remains fixed. 
In accordance with the numerical examples presented in Section~\ref{sec: numerical examples}, we restrict ourselves to two-dimensional cases with $d=2$. 
The shape variation is expressed by a vector $\bm{\theta}\in W^{1,\infty} (\Omega_{\mathrm{all}}; \mathbb{R}^2)$ such that $\bm{\theta}\cdot \nvec=0$ on $\partial \Omega_{\mathrm{all}}$. 
The shape derivative of $J$, denoted by $DJ\cdot \bm{\theta}$, is then defined as \cite{allaire_dapogny_delgado_michailidis_2014},
\begin{align}
     J((\mathrm{Id} + \bm{\theta})(\Omega_a)) = J(\Omega_a)+ DJ\cdot \bm{\theta} + o(\bm{\theta}),
\end{align}
with $\lim_{\bm{\theta\to 0}} \frac{|o(\bm{\theta})|}{|| \bm{\theta} ||_{W^{1,\infty}}}= 0$.
$J((\mathrm{Id} + \bm{\theta})(\Omega_a))$ denotes the objective function with the shape variation introduced by the vector $\bm{\theta}$, while $J(\Omega_a)$ represents the objective function without any shape variation.

Corresponding to an objective function that we define in Section~\ref{sec: numerical examples}, 
we derive the shape derivative for $J$ with the following form:
\begin{align}
    J(\pd,\uvec) = \int_{\Gamma_{\mathrm{meas}}}f_1(\pd)d\Gamma + \int_{\Omega_{\mathrm{all}}}f_2(\pd)h(\uvec,\nabla \uvec)d\Omega, \label{eq: general form of J}
\end{align}
where $\Gamma_{\mathrm{meas}}\subset (\partial \Omega_b\setminus\Gamma)$, and 
$f_1$, $f_2$, and $h$ are real-valued smooth functions.

To derive $DJ\cdot \bm{\theta}$ for $J$ in Eq.~(\ref{eq: general form of J}), 
we follow C\'ea's method \cite{Cea1986} and formulate the Lagrangian based on the boundary value problems for $\pd$ and $\uvec$. 
The boundary value problems for them are expressed as follows, corresponding to the weak forms of Eq.~(\ref{eq: Weak form of p' after transformation}) and Eq.~(\ref{eq: elasticity_Ersatz}):

\begin{align}
    &\left\{ -(\bm{F}^{-T}(\uvecm)\nabla) \cdot \left(\frac{1}{\rhom} \bm{F}^{-T}(\uvecm) \nabla \pdm\right) \right.\nonumber\\
    &\quad\left.-\frac{\omega^2}{\Km}\pdm
    \right\}\det \bm{F}(\uvecm) =0~~~\mathrm{in~}\Omega_m,\\
    &\pda = \pdb~~~\mathrm{on~}\Gamma,\\
    &(\bm{F}^{-T}(\uveca)\nveca) \cdot \left(\frac{1}{\rhoa} \bm{F}^{-T}(\uveca) \nabla \pda\right)\det \bm{F}(\uveca)\nonumber\\ 
    &+ (\bm{F}^{-T}(\uvecb)\nvecb) \cdot \left(\frac{1}{\rhob} \bm{F}^{-T}(\uvecb) \nabla \pdb\right)\det \bm{F}(\uvecb)
    =0~~~\mathrm{on~}\Gamma,\\
    &\nvecb \cdot \left(\frac{1}{\rhob} \nabla \pdb\right) + \frac{i k_b}{\rho_b}\pdb= \left( 1- \frac{\bm{k}\cdot\nvec}{k_b}\right)\frac{ik_b}{\rho_b}P_{\mathrm{in}}~~~\mathrm{on~}\Gamma_{\mathrm{in}},\\
    &\nvecb \cdot \left(\frac{1}{\rhob} \nabla \pdb\right) + \frac{i k_b}{\rho_b}\pdb= 0~~~\mathrm{on~}\Gamma_{\mathrm{out}},\\
    &\nvecb \cdot \left(\frac{1}{\rhob} \nabla \pdb\right)= 0~~~\mathrm{on~}\partial \Omega_b \setminus(\Gamma\cup\Gamma_{\mathrm{in}}\cup \Gamma_{\mathrm{out}}),\\
    &(\bm{F}^{-T}(\uveca)\nveca) \cdot \left(\frac{1}{\rhoa} \bm{F}^{-T}(\uveca) \nabla \pda\right)\det \bm{F}(\uveca) = 0~~~\mathrm{on~}\partial\Omega_a\setminus\Gamma,
\end{align}
\begin{align}
    -\nabla\cdot \sigmam(\uvecm) &= 0~~~\mathrm{in~}\Omega_m,\\
    \uveca &= \uvecb~~~\mathrm{on~}\Gamma,\\
    \sigmaa(\uveca)\cdot\nveca + \sigmab(\uvecb)\cdot\nvecb &= 0  ~~~\mathrm{on~}\Gamma,\\
    \sigmaa(\uveca)\cdot\nveca &= \bm{g}~~~\mathrm{on~}\Gamma_t,\\
    \sigmaa(\uveca)\cdot\nveca &= 0~~~\mathrm{on~}\partial \Omega_a \setminus(\Gamma\cup \Gamma_u\cup\Gamma_t),\\
    \uveca &= 0~~~\mathrm{on~}\Gamma_u,\\
    \uvecb &= 0~~~\mathrm{on~}\partial\Omega_b\setminus\Gamma,
\end{align}
where the index $m=a,b$ is used to represent each domain, $\Omega_a$ and $\Omega_b$, respectively. 
Quantities denoted with $(m)$ correspond to those in the domain $\Omega_m$.
In accordance with C\'ea's method \cite{Cea1986}, the Lagrangian can be formulated as follows:
\begin{align}
    &L(\bm{\theta},\hatp,\hatq,\hatu,\hatv)  \nonumber\\
    &= J(\hatp,\hatu)\nonumber\\
    &+\mathrm{Re}\left[ \sum_{m=a,b}\int_{(\mathrm{Id}+\bm{\theta})\Omega_m}\left\{ \frac{1}{\rhom} (\bm{F^{-T}}(\hatum) \nabla \hatpm)\cdot (\bm{F^{-T}}(\hatum) \nabla \hatqm) \right.\right.\nonumber\\
    &\quad - \left.\frac{\omega^2}{\Km}\hatpm \hatqm
    \right\} \det \bm{F}(\hatum) d\Omega\nonumber\\
    &-\frac{1}{2}\int_{(\mathrm{Id}+\bm{\theta})\Gamma}(\hatpa - \hatpb)
    \left\{ \det \bm{F}(\hatua) ( \bm{F^{-T}}(\hatua)\nveca)\cdot \left(\frac{1}{\rhoa}\bm{F^{-T}}(\hatua)\nabla \hatqa\right) \right. \nonumber\\
    &\quad-\det \bm{F}(\hatub)(\bm{F^{-T}}(\hatub)\nvecb)\cdot \left(\frac{1}{\rhob}\bm{F^{-T}}(\hatub)\nabla \hatqb\right)
    \left. \right\}
    d\Gamma\nonumber\\
    &-\frac{1}{2}\int_{(\mathrm{Id}+\bm{\theta})\Gamma}(\hatqa - \hatqb)
    \left\{ \det \bm{F}(\hatua) ( \bm{F^{-T}}(\hatua)\nveca)\cdot \left(\frac{1}{\rhoa}\bm{F^{-T}}(\hatua)\nabla \hatpa\right) \right. \nonumber\\
    &\quad-\det \bm{F}(\hatub)(\bm{F^{-T}}(\hatub)\nvecb)\cdot \left(\frac{1}{\rhob}\bm{F^{-T}}(\hatub)\nabla \hatpb\right)
    \left. \right\}
    d\Gamma\nonumber\\
    &+\int_{\Gamma_\mathrm{in}}\left\{\frac{ik_b}{\rho_b}\hatp -\left( 1- \frac{\bm{k}\cdot\nvecb}{k_b}\right)\frac{ik_b}{\rho_b}P_{\mathrm{in}}\right\}\hatq d\Gamma
    +\int_{\Gamma_\mathrm{out}}\frac{ik_b}{\rho_b}\hatp \hatq d\Gamma
    \nonumber\\    &+\sum_{m=a,b}\int_{(\mathrm{Id}+\bm{\theta})\Omega_m}\sigmam(\hatum):\varepsilon(\hatvm)d\Omega\nonumber\\
    &-\frac{1}{2}\int_{(\mathrm{Id}+\bm{\theta})\Gamma}(\hatua - \hatub)\cdot 
    \{\sigmaa(\hatva)\cdot \nveca - \sigmab(\hatvb)\cdot \nvecb\}d\Gamma\nonumber\\
    &-\frac{1}{2}\int_{(\mathrm{Id}+\bm{\theta})\Gamma}(\hatva - \hatvb)\cdot 
    \{\sigmaa(\hatua)\cdot \nveca - \sigmab(\hatub)\cdot \nvecb\}d\Gamma\nonumber\\
    &-\int_{\Gamma_t}\bm{g}\cdot \hatva d\Gamma
    -\int_{\Gamma_u}\left\{ (\sigmaa(\hatua)\cdot \nveca)\cdot \hatva +(\sigmaa(\hatva)\cdot \nveca)\cdot \hatua
    \right\}d\Gamma\nonumber\\
    &\left.-\int_{\partial \Omega_b \setminus \Gamma}\left\{ (\sigmab(\hatub)\cdot \nvecb)\cdot \hatvb +(\sigmab(\hatvb)\cdot \nvecb)\cdot \hatub
    \right\}d\Gamma  \right],
\end{align}
where the boundary integrals over the interface $\Gamma$ represent transmission conditions. 

At the optimal point of the Lagrangian $L$, the following optimality conditions hold:
\begin{align}
    &\left.\left<\frac{\partial L}{\partial \hatp},\psi \right>\right|_{\mathrm{opt}} = 0,
    \label{eq: opt_cond_p}\\
    &\left.\left<\frac{\partial L}{\partial \hatq},\psi \right>\right|_{\mathrm{opt}} = 0,
    \label{eq: opt_cond_q}\\
    &\left.\left<\frac{\partial L}{\partial \hatu},\bm{\eta} \right>\right|_{\mathrm{opt}} = 0,
    \label{eq: opt_cond_u}\\
    &\left.\left<\frac{\partial L}{\partial \hatv},\bm{\eta} \right>\right|_{\mathrm{opt}} = 0,
    \label{eq: opt_cond_v}
\end{align}
where the brackets represent the partial derivative in the direction of the second argument. ``opt'' indicates the optimal point of the Lagrangian.
The optimality conditions in Eq.~(\ref{eq: opt_cond_q}) and Eq.~(\ref{eq: opt_cond_v}) imply that $\hatp=\pd(\Omega_a)$ and $\hatu = \uvec(\Omega_a)$ at the optimal point of $L$. 
On the other hand, the optimality conditions in Eq.~(\ref{eq: opt_cond_p}) and Eq.~(\ref{eq: opt_cond_u}) are fulfilled if  $\hatq=\qd(\Omega_a)$ and $\hatv = \vvec(\Omega_a)$ satisfy the following adjoint equations at the optimal point of $L$:
\begin{align}
    &\int_{\Omega}\left\{ \frac{1}{\rho(\bm{x})}
    (\bm{F^{-T}}(\uvec)\nabla \psi) \cdot (\bm{F^{-T}}(\uvec)\nabla \qd) - \frac{\omega^2}{K(\bm{x})}\psi \qd \right\}\det \bm{F}(\uvec) d\Omega \nonumber\\
    &= -\int_{\Gamma_{\mathrm{meas}}}\frac{\partial f_1}{\partial p}\psi d\Gamma
    -\int_{\Omega_{\mathrm{all}}}\frac{\partial f_2}{\partial p}h(\uvec,\nabla \uvec)  \psi d\Omega~~~\forall \psi \in H^1(\Omega_{\mathrm{all}}), \label{eq: adj_q}
\end{align}

\begin{align}
    &\int_{\Omega}\sigma(\bm{\eta}):\varepsilon(\vvec) d\Omega\nonumber\\
    &=-\int_{\Omega_{\mathrm{all}}} f_2(p)\left\{ \frac{\partial h}{\partial \uvec}\cdot \bm{\eta}
    + \frac{\partial h}{\partial \nabla \uvec}: \nabla \bm{\eta}
    \right\}d\Omega
    \nonumber\\
    &-\int_{\Omega}\left( \frac{\partial \det \bm{F}(\uvec)}{\partial \uvec}:\nabla \bm{\eta}\right)
    \left\{ \frac{1}{\rho(\bm{x})}
    (\bm{F^{-T}}(\uvec)\nabla \pd) \cdot (\bm{F^{-T}}(\uvec)\nabla \qd) - \frac{\omega^2}{K(\bm{x})} \pd\qd \right\}d\Omega\nonumber\\
    &-\int_{\Omega}\det \bm{F}(\uvec)\frac{1}{\rho(\bm{x})}\left\{
    \left(\frac{\partial \bm{F^{-T}}(\uvec)}{\partial \uvec}:\nabla  \bm{\eta}\right)\nabla\pd 
    \right\}\cdot (\bm{F^{-T}}(\uvec)\nabla \qd) d\Omega\nonumber\\
    &-\int_{\Omega}\det \bm{F}(\uvec)\frac{1}{\rho(\bm{x})}
    (\bm{F^{-T}}(\uvec)\nabla \pd)
    \cdot \left\{\left(\frac{\partial \bm{F^{-T}}(\uvec)}{\partial \uvec}:\nabla  \bm{\eta}\right)\nabla \qd\right\}d\Omega
    ~~~{\forall \bm{\eta} \in W(\Omega_{\mathrm{all}})}.\label{eq: adj_v}
\end{align}

Using these adjoint variables, the shape derivative is derived as follows \cite{allaire_dapogny_delgado_michailidis_2014,allaire2011damage}:
\begin{align}
    DJ\cdot \bm{\theta}&=\left<\frac{\partial L(0,\pd,\qd,\uvec,\vvec)}{\partial \bm{\theta}},\bm{\theta} \right>\nonumber\\
    &=\mathrm{Re}\left[
    \sum_{m=a,b}\int_{\Gamma} \left\{ \frac{1}{\rhom} (\bm{F^{-T}}(\uvecm) \nabla \pdm)\cdot (\bm{F^{-T}}(\uvecm) \nabla \right.\qdm) \right.\nonumber\\ 
    &\left.- \frac{\omega^2}{\Km}\pdm \qdm
    \right\} \det \bm{F}(\uvecm) (\bm{\theta}\cdot \nvecm)d\Gamma\nonumber\\
    &-\frac{1}{2}\int_{\Gamma}(\nabla \pda \cdot \nvec - \nabla \pdb \cdot \nvec)\left\{ \det \bm{F}(\uveca) ( \bm{F^{-T}}(\uveca)\nveca)\cdot \left(\frac{1}{\rhoa}\bm{F^{-T}}(\uveca)\nabla \qda\right) \right. \nonumber\\
    &-\det \bm{F}(\uvecb)(\bm{F^{-T}}(\uvecb)\nvecb)\cdot \left(\frac{1}{\rhob}\bm{F^{-T}}(\uvecb)\nabla \qdb\right)
    \left. \right\}(\bm{\theta}\cdot \nvec)
    d\Gamma\nonumber\\
    &-\frac{1}{2}\int_{\Gamma}(\nabla \qda \cdot \nvec - \nabla \qdb \cdot \nvec)\left\{ \det \bm{F}(\uveca) ( \bm{F^{-T}}(\uveca)\nveca)\cdot \left(\frac{1}{\rhoa}\bm{F^{-T}}(\uveca)\nabla \pda\right) \right. \nonumber\\
    &-\det \bm{F}(\uvecb)(\bm{F^{-T}}(\uvecb)\nvecb)\cdot \left(\frac{1}{\rhob}\bm{F^{-T}}(\uvecb)\nabla \pdb\right)
    \left. \right\}(\bm{\theta}\cdot \nvec)
    d\Gamma\nonumber\\
    &+\sum_{m=a,b}\int_{\Gamma}\sigmam(\uvecm):\varepsilon(\vvecm)(\bm{\theta}\cdot \nvecm)d\Gamma\nonumber\\
    &-\frac{1}{2}\int_{\Gamma}(\nabla\uveca\cdot\nvec - \nabla\uvecb\cdot \nvec)\cdot 
    \{\sigmaa(\vveca)\cdot \nveca - \sigmab(\vvecb)\cdot \nvecb\}(\bm{\theta}\cdot\nvec)d\Gamma\nonumber\\
    &\left.-\frac{1}{2}\int_{\Gamma}(\nabla\vveca\cdot\nvec - \nabla\vvecb\cdot \nvec)\cdot 
    \{\sigmaa(\uveca)\cdot \nveca - \sigmab(\uvecb)\cdot \nvecb\}(\bm{\theta}\cdot\nvec)d\Gamma \right],
    \label{eq: derived_shape_derivative_orig}
\end{align}
where we set $\nvec = \nveca = -\nvecb$ on $\Gamma$. 

By utilizing the transmission conditions on $\Gamma$, the shape derivative can be expressed in terms of the jump on $\Gamma$, as shown below:
\begin{align}
    &DJ\cdot \bm{\theta} = \int_{\Gamma} \jump{V^{(m)}} (\bm{\theta}\cdot \bm{n})d\Gamma, \label{eq: Vm}
\end{align}
where $\jump$ denotes the jump on $\Gamma$, i.e., $\jump{c^{(m)}}= c^{(a)}-c^{(b)}$. 
The function $V^{(m)}$ is defined as follows:  
\begin{align}
    &V^{(m)} = \mathrm{Re}\left[
    \left\{ \frac{1}{\rhom} (\bm{F^{-T}}(\uvecm) \nabla \pdm)\cdot (\bm{F^{-T}}(\uvecm) \nabla \right.\qdm)\left.- \frac{\omega^2}{\Km}\pdm \qdm
    \right\} \det \bm{F}(\uvecm)  \right. \nonumber\\
    &-(\nabla \pdm \cdot \nvec)\left\{\det \bm{F}(\uvecm) ( \bm{F^{-T}}(\uvecm)\nvec)\cdot \left(\frac{1}{\rhom}\bm{F^{-T}}(\uvecm)\nabla \qdm\right)\right\}\nonumber\\
    &-(\nabla \qdm \cdot \nvec)\left\{\det \bm{F}(\uvecm) ( \bm{F^{-T}}(\uvecm)\nvec)\cdot \left(\frac{1}{\rhom}\bm{F^{-T}}(\uvecm)\nabla \pdm\right)\right\}\nonumber\\
    &+\sigmam(\uvecm):\varepsilon(\vvecm)\nonumber\\
    &\left.-(\nabla\uvecm\cdot\nvec)\cdot\{\sigmam(\vvecm)\cdot \nvec \}-(\nabla\vvecm\cdot\nvec)\cdot\{\sigmam(\uvecm)\cdot \nvec \}
    \right].
    \label{eq: Vm}
\end{align}

It should be noted that the shape derivative of an objective function $J(p)$, which depends on the acoustic pressure $p$ when the structure is not deformed, can be obtained by setting $\uvec = \bm{0}$ in Eq.~(\ref{eq: derived_shape_derivative_orig}).

\section{Level set-based shape and topology optimization method}\label{sec: Level set-based topology optimization}
To address the optimization problem defined in Eq.~(\ref{eq: topopt}), a level set-based shape and topology optimization \cite{allaire2004structural} is introduced. 
Specifically, the method proposed by Feppon et al. \cite{feppon2017introducing} is adopted in this work. 
The approach is briefly presented in this section. 

The method utilizes a level set function $\phi$ to represent each material phase, following the standard level set method. 
The level set function is defined to delineate the structural domain $\Omega_a$, the acoustic domain $\Omega_b$, and the interface $\Gamma$, as follows:
\begin{eqnarray}
	\left\{
	\begin{array}{ll}
		\phi(\bm{x})>0 &\mathrm{if}~~\bm{x}\in \Omega_a,\\
		\phi(\bm{x})= 0 &\mathrm{if}~~\bm{x}\in \Gamma,\\
		\phi(\bm{x})< 0 &\mathrm{if}~~\bm{x}\in \Omega_b.\label{eq:profile of LSF}
	\end{array}
	\right. 
\end{eqnarray}

To regularize the optimization problem, 
the level set-based optimization method proposed in \cite{feppon2017introducing} employs perimeter penalization. 
We modify the objective function $J$ to incorporate the perimeter control, which is achieved through a modified objective function $J_\tau$ defined as:
\begin{align}
    J_\tau = J + \tau |\Gamma|, \label{eq: opt_general_lsf}
\end{align}
where $|\Gamma|$ denotes the perimeter of the interface $\Gamma$, given by $|\Gamma|=\int_{\Gamma}d\Gamma$, and $\tau$ is a positive parameter serving as a weighting factor for controlling the perimeter.
The shape derivative of the modified objective function $J_\tau$ can be estimated as \cite{feppon2017introducing}:
\begin{align}
 D J_\tau \cdot \bm{\theta} &= D J \cdot \bm{\theta} +  \tau\int_{\Gamma}\kappa (\bm{\theta}\cdot \nvec)d\Gamma\nonumber\\
&=\int_{\Gamma}(\jump{V^{(m)}} + \tau \kappa  )(\bm{\theta}\cdot \nvec) d\Gamma,
\label{eq: shape derivative Jtau}
\end{align}
where $\nvec=\nveca$, and 
$\kappa$ denotes the mean curvature of the interface. 
By examining Eq.~(\ref{eq: shape derivative Jtau}), 
we can determine the descent direction to minimize $J_\tau$ as $\bm{\theta} = -(\jump{V^{(m)}} + \tau \kappa  )\nvec$.
To minimize the objective function in Eq.~(\ref{eq: opt_general_lsf}), the domain $\Omega_a$ can be updated by solving the Hamilton--Jacobi equation for $\phi$ given by: 
\begin{align}
    \frac{\partial \phi}{\partial t} - v |\nabla \phi| = 0,
    \label{eq: Hamilton-Jacobi}
\end{align}
where $t$ represents a fictitious time, and $v=-(\jump{V^{(m)}} + \tau \kappa)$ denotes the descent direction. 
The negative sign in front of $v$ is a consequence of the definition of $\phi$ in Eq.~(\ref{eq:profile of LSF}), where $\phi$ is positive in $\Omega_a$. 

To update the domain $\Omega_a$, a different updating scheme based on the profile of the level set function is introduced in \cite{feppon2017introducing}, rather than using Eq.~(\ref{eq: Hamilton-Jacobi}).
Specifically, a level set function satisfying Eq.~(\ref{eq:profile of LSF}) is introduced in \cite{feppon2017introducing} as follows: 
\begin{align}
    \phi = 2\tilde{H}(-d_{\Omega_a}) - 1,
    \label{eq: LSF-distanceFunc}
\end{align}
where $d_{\Omega_a}$ is the signed-distance function for $\Omega_a$, taking negative values in $\Omega_a$ and positive values in $\Omega_b$.
$\tilde{H}$ is the smoothed Heaviside function with a parameter $\epsilon$, defined as follows:
\begin{equation}
	\tilde{H}(t)=
	\begin{cases}
		0 &  {\rm if} \ t<-\epsilon,						 \\
		\frac{1}{2}+\frac{t}{2\epsilon}+\frac{1}{2\pi}\sin (\frac{\pi t}{\epsilon}) &  {\rm if} \ |t|\le \epsilon, \\
            1 &  {\rm if} \ t> \epsilon.
	\end{cases}										\label{eq: Hev}
\end{equation}
Using the profile of $\phi$ given in Eq.~(\ref{eq: LSF-distanceFunc}), 
it follows that $\nabla \phi = -\frac{2}{\epsilon}\nvec$ and $\nabla^2 \phi = -\frac{2}{\epsilon}\kappa$ on the interface $\Gamma$, where $\nvec$ and $\kappa$ are the normal vector and mean curvature, respectively.
Therefore, Eq.~(\ref{eq: Hamilton-Jacobi}) can be replaced by the following equation:
\begin{align}
     \frac{\partial \phi}{\partial t}=-\frac{2}{\epsilon} \jump{V^{(m)}} + \tau \nabla^2 \phi,
    \label{eq: reaction-diffusion eq}
\end{align}
Eq.~(\ref{eq: reaction-diffusion eq}) can be considered a reaction--diffusion equation by extending $\jump{V^{(m)}}$ to the entire region. 
The reaction term on the right-hand side of Eq.~(\ref{eq: reaction-diffusion eq}) minimizes the objective function $J$, while the diffusion term minimizes $\tau |\Gamma|$. 
A larger value of $\tau$ leads to a simpler structure with less perimeter \cite{feppon2017introducing}, as in a level set-based topology optimization method proposed by Yamada et al. \cite{yamada2010topology}.
To ensure that the minimization of $J$ is undisturbed, we use a small value of $\tau$. 

In addition to guaranteeing a simpler structure with less perimeter, 
the use of Eq.~(\ref{eq: reaction-diffusion eq}) instead of Eq.~(\ref{eq: Hamilton-Jacobi}) is beneficial in terms of numerical treatment. 
Numerically solving Eq.~(\ref{eq: Hamilton-Jacobi}) using the FEM with an unstructured mesh requires a special treatment, such as introducing the method of characteristics \cite{allaire2014shape}. 
However, the reaction--diffusion equation in Eq.~(\ref{eq: reaction-diffusion eq}) can be solved using the standard FEM solver by introducing the finite difference method in the direction of $t$. A more detailed description of the discretization of Eq.~(\ref{eq: reaction-diffusion eq}) is provided in Section~\ref{sec: Discretization RDE}. 

\section{Numerical implementation}\label{sec:Numerical implementation}

\subsection{Numerical procedure to obtain the signed-distance function}
In order to maintain the profile of the level set function in Eq.~(\ref{eq: LSF-distanceFunc}), it is necessary to estimate the signed-distance function $d_{\Omega_a}$ for $\Omega_a$ during the optimization procedure. 
To achieve this, we introduce a method for obtaining an approximate solution of the Eikonal equation \cite{churbanov2019numerical}.
Specifically, we introduce an auxiliary boundary value problem for a scalar field $v_e$ given by:
\begin{align}
-\alpha_e^2\nabla^2 v_e + v_e &= 0~~~\mathrm{in~}\Omega,\nonumber\\
v_e &= 1~~~\mathrm{on~}\Gamma,\label{eq: PDE of approximiated eikonal}
\end{align}
where $\alpha_e$ is a positive constant. 
By using the solution of Eq.~(\ref{eq: PDE of approximiated eikonal}), 
a scalar-valued function $u_e$ is defined as follows:
\begin{align}
	u_e = -\alpha_e \log(v_e)(1-2\chi_{\Omega_a}),\label{eq: u_e}
\end{align}
where $\chi_{\Omega_a}$ is the characteristic function of $\Omega_a$. 
It has been proven that 
as $\alpha_e$ approaches 0, $u_e$ approaches the signed distance function $d_{\Omega_a}$ \cite{churbanov2019numerical}. 
However, to avoid numerical instability for solving the boundary value problem in Eq.~(\ref{eq: PDE of approximiated eikonal}), $\alpha_e$ should be set to the same order as the element size in the FEM analysis. The detailed setting of $\alpha_e$ is provided in Section~\ref{sec: numerical examples}.

\subsection{Extension of the velocity field}
To extend the velocity field $\jump{V^{(m)}}$ in Eq.~(\ref{eq: reaction-diffusion eq}) to the entire domain $\Omega_{\mathrm{all}}$, we introduce the following equation \cite{allaire2011damage}: 
\begin{align}
    \int_{\Omega_{\mathrm{all}}} (\alpha_{\nu}^2 \nabla \nu \cdot \nabla \tilde{\nu} + \nu \tilde{\nu})d\Omega 
    = \int_{\Gamma}\jump{V^{(m)}}\tilde{\nu}d\Gamma~~~\forall \tilde{\nu}\in H^1(\Omega_{\mathrm{all}}),
    \label{eq: extension_orig}
\end{align}
where $\alpha_\nu$ is a positive small parameter that is set to the order of the element size. 
The solution $\nu \in H^1(\Omega_{\mathrm{all}})$ is the extended velocity in the entire region.

It is difficult to numerically estimate the jumped quantities on $\Gamma$; thus, we replace the right-hand side of Eq.~(\ref{eq: extension_orig}) with volume integrals by introducing an approximate delta function $\delta(u_e)$. 
Specifically, we use the solution of the following equation instead of Eq.~(\ref{eq: extension_orig}):
\begin{align}
    \int_{\Omega_{\mathrm{all}}} (\alpha_{\nu}^2\nabla \nu \cdot \nabla \tilde{\nu} + \nu \tilde{\nu})d\Omega 
    = \int_{\Omega_a}V^{(a)} \delta(u_e)\tilde{\nu}d\Omega - \int_{\Omega_b}V^{(b)} \delta(u_e)\tilde{\nu}d\Omega
    ~~~\forall \tilde{\nu}\in H^1(\Omega_{\mathrm{all}}).
    \label{eq: extention_mod}
\end{align}
Here, $\delta(u_e) = \tilde{H}'(-u_e)$ approximates the delta function on $\Gamma$, and the normal vector $\nvec$ in the formula for $V_m$ (Eq.~(\ref{eq: Vm})) is approximated by $\nvec\approx \nabla u_e / \sqrt{|\nabla u_e|^2  + \epsilon_0^2 }$ using a small value of $\epsilon_0$, which we fixed at $1\times 10^{-3}$.

Solving Eq.~(\ref{eq: extention_mod}) via the FEM enables us to obtain the extended velocity field $\nu$ for updating $\phi$ with the reaction--diffusion equation presented in Eq.~(\ref{eq: reaction-diffusion eq}).

\subsection{Discretization of the reaction--diffusion equation}\label{sec: Discretization RDE}
As previously mentioned, the level set function is updated using Eq.~(\ref{eq: reaction-diffusion eq}). 
For practical purposes, a design domain $D \subset \Omega_{\mathrm{all}}$ is defined, within which Eq.~(\ref{eq: reaction-diffusion eq}) is solved.  
The remaining region outside $D$ is considered as non-design domains, where the material phase remains unchanged during the optimization process. 
Thus, the structural design is optimized only within $D$.

To discretize Eq.~(\ref{eq: reaction-diffusion eq}) over $D$, 
we introduce the finite difference method for the fictitious time $t$, while using the FEM for space. 
Firstly, we discretize $t$ with a time step $\Delta t$. We denote the discretized $\phi$ with respect to $t$ as $\phi^{(n)}$, such that $\phi^{(n)} = \phi(n \Delta t)$,  where $n\ge 0$ is an integer value. 
Using the implicit scheme and the extended velocity field $\nu$ obtained by solving Eq.~(\ref{eq: extention_mod}), we discretize Eq.~(\ref{eq: reaction-diffusion eq}) as
\begin{align}
    \frac{\phi^{(n+1)}-\phi^{(n)}}{\Delta t} = -\frac{2}{\epsilon}\nu + \tau \nabla^2 \phi^{(n+1)}~~~\mathrm{in~}D.\label{eq: RDE_dimensional}
\end{align}
We use a non-dimensionalized form of the reaction--diffusion equation instead of Eq.~(\ref{eq: RDE_dimensional}) and prepare boundary conditions for the level set function to solve it using the FEM. In this research, we consider the following boundary value problem for $\phi^{(n+1)}$:
\begin{align}
    \frac{\phi^{(n+1)}-\phi^{(n)}}{\Delta t} &= -C\overline{\nu} + \tau L^2 \nabla^2 \phi^{(n+1)}~~~\mathrm{in~}D,\nonumber\\
    \phi^{(n+1)} &= 1~~~\mathrm{on~}\Gamma_{\phi p},\nonumber\\
    \phi^{(n+1)} &= -1~~~\mathrm{on~}\Gamma_{\phi n},\nonumber\\
    \nvec \cdot \nabla \phi^{(n+1)} &= 0 ~~~ \mathrm{on~}\partial D \setminus (\Gamma_{\phi p} \cup \Gamma_{\phi n}),\label{eq: RDE_non-dimensional}
\end{align}
where $L$ is the characteristic length of the design domain $D$. 
$\overline{\nu}$ represents the normalized extended velocity field, defined as follows:
\begin{align}
    \overline{\nu} = \nu \left( \frac{ \int_{D} |\nu \delta(u_e)|d\Omega  }{\int_{D} |\delta(u_e)|d\Omega }   \right)^{-1} .
\end{align}
$C$ is a positive constant to control the magnitude of the velocity. 
For simplicity, we set $C=1$ in the numerical examples, although in \cite{feppon2017introducing}, $C$ is defined with respect to the parameter $\epsilon$, which corresponds to the coefficient of the right-hand side of Eq.~(\ref{eq: reaction-diffusion eq}).
The second and third lines in Eq.~(\ref{eq: RDE_non-dimensional}) represent the Dirichlet boundary conditions
that fix the value of $\phi^{(n+1)}$. 
Specifically, $\Gamma_{\phi p} \subset \partial D$ is a boundary that sets the value of $\phi$ to $1$, while $\Gamma_{\phi n} \subset \partial D$ is a boundary that sets the value of $\phi$ to $-1$. 
In the numerical examples presented in Section~\ref{sec: numerical examples}, we set $\Gamma_{\phi p}$ as boundaries that touch a non-design domain filled with an elastic medium, and $\Gamma_{\phi n}$ as boundaries that touch a non-design domain filled with air. 
The last line in Eq.~(\ref{eq: RDE_non-dimensional}) represents the Neumann boundary condition for $\phi^{(n+1)}$. However, other choices of boundary conditions are also possible. 

Based on Eq.~(\ref{eq: RDE_non-dimensional}), we formulate the following weak form:
\begin{align}
    &\int_{D} \left(\frac{1}{\Delta t} \phi^{(n+1)} \tilde{\phi} + \tau L^2 \nabla \phi^{(n+1)} \cdot \nabla \tilde{\phi} \right) d\Omega \nonumber\\
    &= \int_{D}\left\{\frac{1}{\Delta t} \phi^{(n)}\tilde{\phi} -C\overline{\nu}\tilde{\phi} \right\}d\Omega
    ~~~\forall \tilde{\phi} \in V_{\phi}(D),\label{eq: RDE_weakform}
\end{align}
where $\tilde{\phi}$ represents a test function, and $V_{\phi}(D) = \{ \tilde{\phi} \in H^1(D) |~\tilde{\phi}=0 \mathrm{~on~}\Gamma_{\phi p}\cup \Gamma_{\phi n} \}$ is the functional space for $\tilde{\phi}$. 
Given an initial level set function $\phi^{(0)}$ that represents an initial design, we iteratively solve Eq.~(\ref{eq: RDE_weakform}) to obtain the optimized shape of $\Omega_a$.



\subsection{Optimization flow}
The optimization procedure can be summarized as follows:
\begin{itemize}
    \item[Step 1:] 
    Set the iteration number $n$ to 0, and discretize the computational region using finite elements. Initialize the variables used in the optimization, including the initial level set function $\phi^{(0)}$.
    \item[Step 2:] 
    Refine the finite elements so that the isosurface of $\phi^{(n)}=0$ fits the interface $\Gamma$.
    \item[Step 3:] 
    Solve the auxiliary boundary value problem for $v_e$ in Eq.~(\ref{eq: PDE of approximiated eikonal}) to obtain the approximated signed-distance function $u_e$ in Eq.~(\ref{eq: u_e}). Modify $\phi^{(n)}$ using the formula $\phi^{(n)}=2\tilde{H}(-u_e)-1$, which corresponds to Eq.~(\ref{eq: LSF-distanceFunc}).
    \item[Step 4:] 
    Use the FEM to solve the equations for the state variables (Eq.~(\ref{eq: Weak form of p' after transformation}) and Eq.~(\ref{eq: elasticity_Ersatz})) and estimate the objective function $J$ in Eq.~(\ref{eq: general form of J}). If the value of $J$ has converged, end the optimization procedure. Otherwise, proceed to Step 5.
    \item[Step 5:] 
    Use the FEM to solve the adjoint equations in Eq.~(\ref{eq: adj_q}) and Eq.~(\ref{eq: adj_v}).
    \item[Step 6:] 
    Use the FEM to solve Eq.~(\ref{eq: extention_mod}) and extend the velocity field to the entire region.
    \item[Step 7:] 
    Use the FEM to solve the discretized reaction-diffusion equation in Eq.~(\ref{eq: RDE_weakform}) and obtain the updated level set function $\phi^{(n+1)}$. Increment $n$ by 1 and return to Step 2.
\end{itemize}
All FEM analyses are performed using FreeFEM \cite{MR3043640}, an open-source partial differential equation (PDE) solver. 
In Step 2, the remeshing procedure is implemented using Mmg \cite{dapogny2014three}, an open-source mesher. 
To determine the convergence of the optimization calculation in Step 4, 
the absolute difference between the values of the objective function $J$ for two consecutive iterations is estimated. 
The optimization calculation is terminated if this difference becomes sufficiently small after several optimization iterations.

\section{Numerical examples}\label{sec: numerical examples}
This section presents two types of two-dimensional optimization examples: Case 1 and Case 2. In Case 1, we optimize the surface structure of an open tube to maximize its transmission loss. In Case 2, we optimize the shape of acoustic lenses to maximize the amplitude of reflected waves in a desired region. We conduct optimization calculations with and without deformation to examine its impact. 

In these examples, $\Omega_a$ is assumed to be filled with aluminum, whereas $\Omega_b$ is filled with air. 
The mass density and bulk modulus of aluminum are $\rho_a = 2643~\mathrm{[kg/m^3]}$ and $K_a=6.870\times 10^{10}\mathrm{[Pa]}$, respectively. 
The mass density and bulk modulus of air are $\rho_b = 1.225\mathrm{[kg/m^3]}$ and $K_b=1.426\times 10^{5}\mathrm{[Pa]}$, respectively. 
Poisson's ratio of aluminum is set to be $0.3$. 
The small coefficient $\varepsilon_u$ in Eq.~(\ref{eq: Elasticity tensor}) is set to $0.01$. 

We have validated our numerical analysis for acoustic wave propagation problems with deformation. In Section \ref{sec: Validation}, we compared the acoustic pressure distribution at the optimized designs obtained by the proposed optimization method in Section 2 with a reference method that employs a Lagrangian mesh.

\subsection{Case1: Maximizing transmission loss in a deformable tube}
\begin{figure}[htbp]
	\begin{center}
		\includegraphics[scale=0.25]{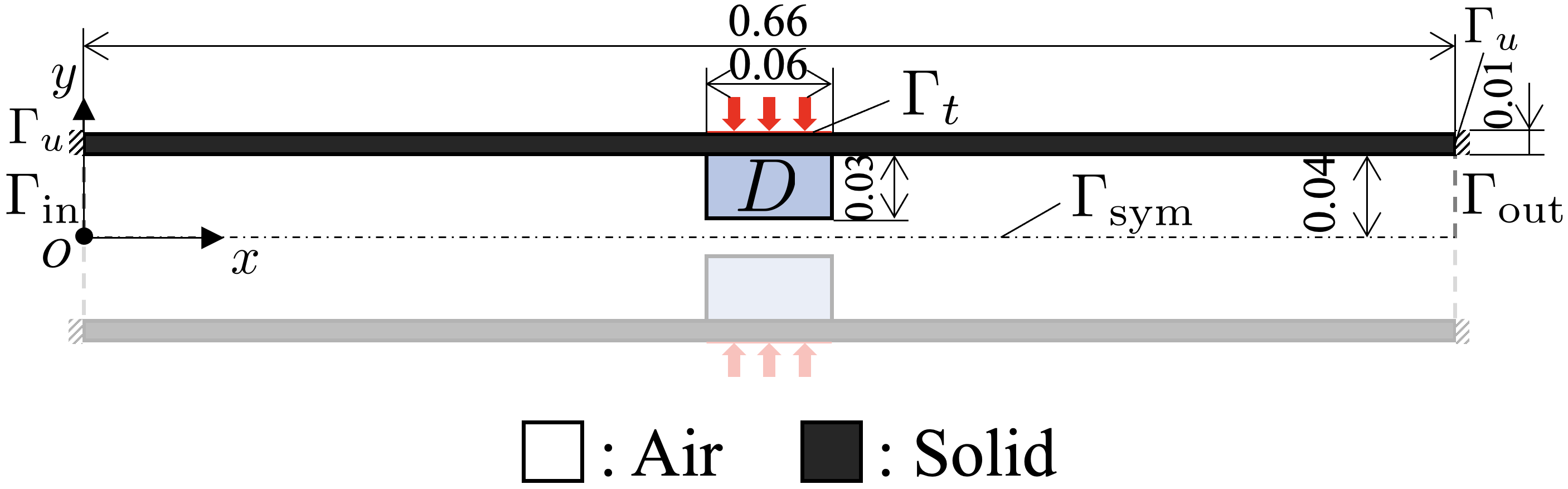}
	\end{center}
	\caption{Geometry setting of Case 1. Units are in [m].}
	\label{fig: geom}
\end{figure}
In this example, we aim to maximize the transmission loss of acoustic waves propagating in an open tube made of elastic material. 
The geometry setting of Case 1 is shown in Figure~\ref{fig: geom}, where the solid-filled and air-filled regions are represented by black and white colors, respectively. 
The blue-colored region denotes the design domain $D$ that will be optimized. 
An incident plane wave propagating along the $x$-direction with a wave vector $\bm{k} = [k_b,0]^T$ and a frequency of 3000~[Hz] is applied at the inlet boundary $\Gamma_{\mathrm{in}}$. 
Fixed displacement constraints are applied on  the boundary $\Gamma_u$, while a traction force $\bm{g}$ is applied on the boundary $\Gamma_t$. 
The boundary conditions and the shape of the design domain $D$ are assumed to be symmetric with respect to the dashed line $\Gamma_{\mathrm{sym}}$, where the symmetric condition is imposed. 
Consequently, only half of the entire region is considered in the finite element analyses.

The objective function $J$ to maximize the transmission loss is expressed as the minimization of the amplitude of $\pd$ on the outlet boundary $\Gamma_{\mathrm{out}}$, as follows: 
\begin{align}
    J = \int_{\Gamma_{\mathrm{out}}} |\pd|^2 d\Gamma. \label{eq: obj in case 1}
\end{align}
We optimize the structure in $D$ to minimize $J$ in Eq.(\ref{eq: obj in case 1}) for three different magnitudes of the traction force, $\bm{g} = [0, g_y]^T$, where $g_y$ is set to $0$, $-1\times 10^8$~[Pa], and $-5\times 10^8~$[Pa], respectively. 
The parameter $\tau$ controlling the perimeter is set to $1\times 10^{-5}$ with a characteristic length of $L=0.06$ [m]. 
The finite elements are refined so that the size of each element in $D$ is around $3.0\times 10^{-3}$~[m]. 
The parameter $\epsilon$ in the smoothed Heaviside function $\tilde{H}$ is set to $\epsilon = 3.0\times 10^{-3}$.
The parameter $\alpha_e$ in Eq.~(\ref{eq: u_e}) is set to $\alpha_e=4.5\times 10^{-3}$. In addition,
$\alpha_\nu= 4.5\times 10^{-3}$ is used to extend the velocity field. 
In these examples, the optimization calculation is halted when the absolute value of the difference between the values of $J$ for two consecutive iterations is smaller than $1\times 10^{-6}$ after 50 iterations of the calculations. 

\begin{figure}[htbp]
	\begin{center}
		\includegraphics[scale=0.5]{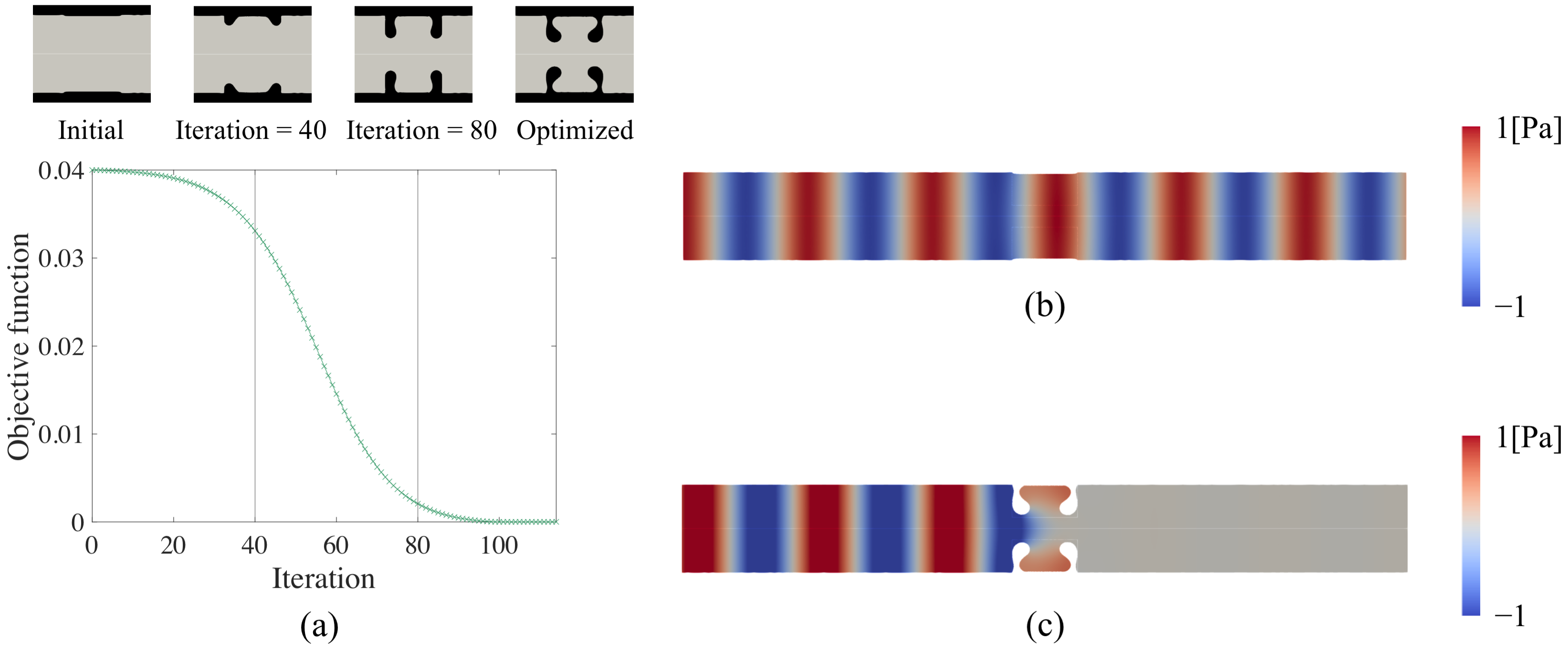}
	\end{center}
	\caption{(a) History of the objective function with structures during the optimization when $g_y=0$. The black region represents the solid region, whereas the gray region represents the air-filled region.
                 (b) Acoustic pressure distribution in the air-filled region with the initial structure. 
                 (c) Acoustic pressure distribution in the air-filled region with the optimal structure. }
	\label{fig: undeform}
\end{figure}
We first present the results for the case without deformation ($g_y=0$). 
Figure~\ref{fig: undeform} illustrates the optimized results. 
Figure~\ref{fig: undeform}(a) displays the history of $J$ with initial, immediate, and optimized structures during the optimization. 
As shown in the upper left of Fig.~\ref{fig: undeform}(a), the initial design was set so that $D$ was filled with air. 
Figure~\ref{fig: undeform}(b) displays the acoustic pressure distribution in the air-filled region with the initial design. 
The incident plane wave propagated through the tube toward the outlet, resulting in almost perfect transmission, with a value of $J$ of 0.04 for the initial design. 
After several iterations of the optimization, the value of $J$ smoothly decreased, and four bumps appeared in $D$. 
The optimization calculation was halted at the 114-th iteration, and the optimized design was obtained, as shown in the upper right of Fig.~\ref{fig: undeform}(a). 
The value of $J$ in the optimized design was $9.77\times 10^{-6}$, significantly smaller than the initial value.
Figure~\ref{fig: undeform}(c) shows the acoustic pressure distribution in the air-filled region with the optimized design. 
The amplitude of the acoustic pressure was reduced due to the reflection of the incident wave on the optimized design in $D$.

\begin{figure}[htbp]
	\begin{center}
		\includegraphics[scale=0.55]{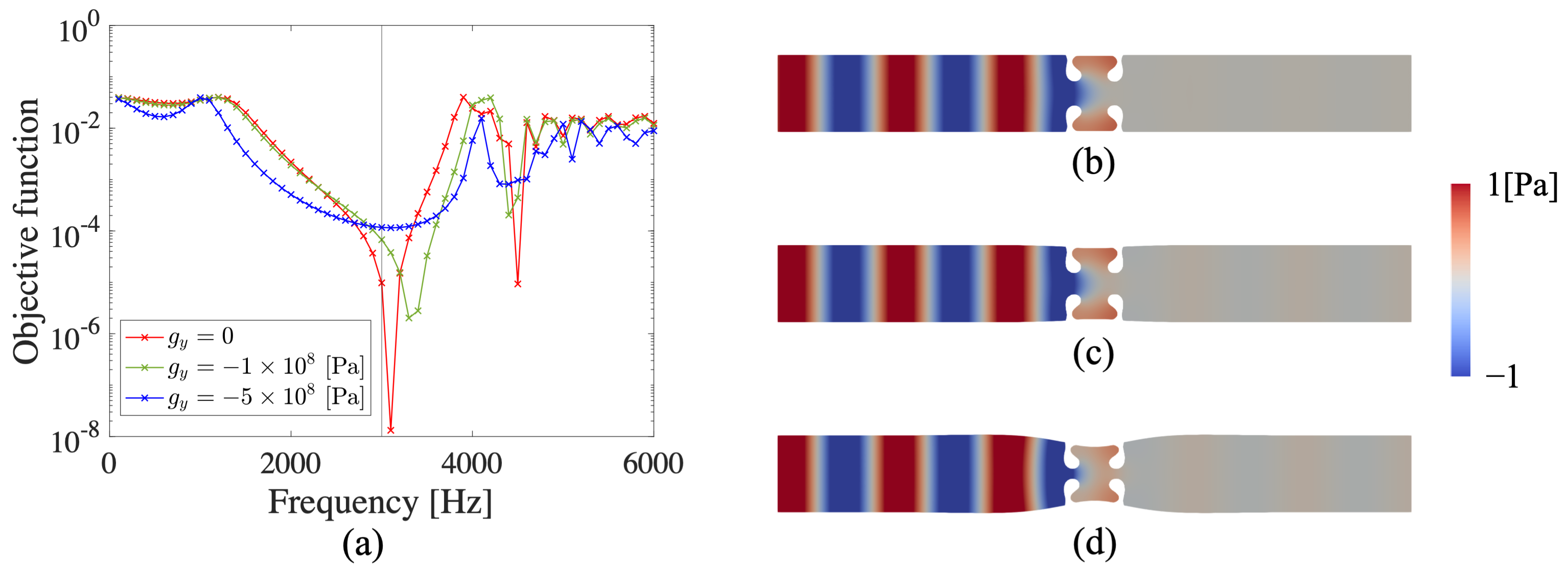}
	\end{center}
	\caption{(a) Frequency responses of $J$ of the optimized design for $g_y=0$. The red, green, and blue plots represent the frequency responses when $g_y=0$, $g_y=-1\times 10^8~$[Pa], and $g_y=-5\times 10^8~$[Pa], respectively. 
    (b)--(d) Acoustic pressure distribution at 3000 [Hz] with the optimized design for $g_y=0$ when (b) $g_y=0$, (c) $g_y=-1\times 10^8~$[Pa], and (d) $g_y=-5\times 10^8~$[Pa].
    }
	\label{fig: Objfreq_optpu}
\end{figure}
Figure~\ref{fig: Objfreq_optpu}(a) shows the frequency responses of the optimized design's $J$ for $g_y=0$. 
The red line corresponds to the case without deformation ($g_y = 0$). 
It is characterized by a sharp curve with a peak frequency around the target frequency, implying a resonance property of the structure. 
In effect, the value of $J$ is almost minimum at 3000~[Hz], which is shown by the vertical line in the figure. 
However, the frequency responses change when a non-zero value of $g_y$ is applied. 
The blue and green lines in Fig.~\ref{fig: Objfreq_optpu}(a) show the frequency curves for the cases with $g_y=-1\times 10^8~$[Pa] and $g_y=-5\times 10^8~$[Pa], respectively. 
The values of $J$ at the target frequency in these plots are larger than those without deformation. 
Figures~\ref{fig: Objfreq_optpu}(b)--(d) show the acoustic pressure distribution at 3000~[Hz] corresponding to these plots. 
More transmitted waves were observed at the outlet in Figs.~\ref{fig: Objfreq_optpu}(c) and (d) compared to Fig.~\ref{fig: Objfreq_optpu}(a). 
Based on these results, the obtained design was found to work without deformation, but not when it deforms.

\begin{figure}[htbp]
	\begin{center}
		\includegraphics[scale=0.5]{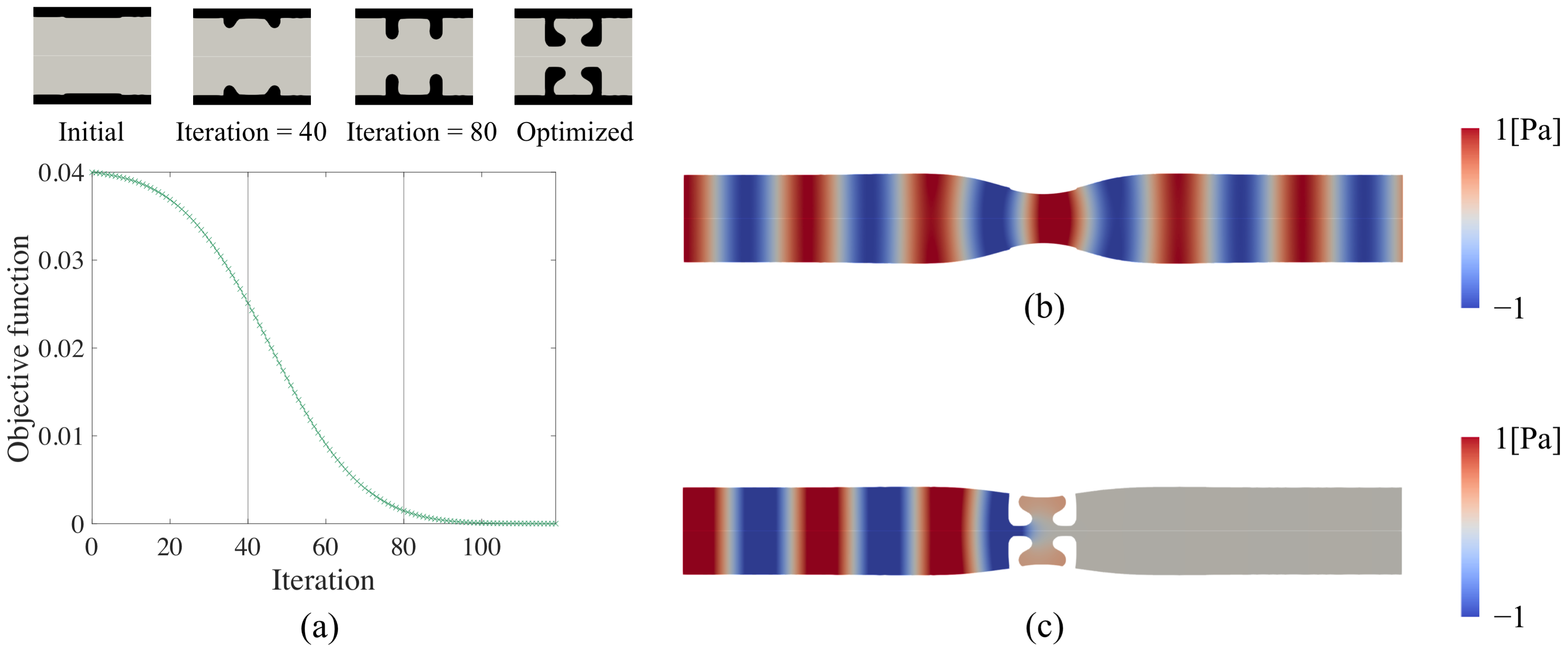}
	\end{center}
	\caption{
                 (a) History of the objective function with structures during the optimization when $g_y=-5\times 10^{8}~\mathrm{[Pa]}$. 
                 (b) Acoustic pressure distribution in the air-filled region with the initial structure. 
                 (c) Acoustic pressure distribution in the air-filled region with the optimal structure. }
	\label{fig: 5e8}
\end{figure}

Next, we present the optimized results with $g_y = -5\times 10^8~$[Pa]. 
In this case, the solid region was deformed due to the presence of the traction force $\bm{g}$, which is different from the previous example. 
Figure~\ref{fig: 5e8} illustrates the optimized results. 
As shown in Fig.~\ref{fig: 5e8}(a), we used the same initial design as in Fig.~\ref{fig: undeform}. 
The value of $J$ with the initial design was 0.04. 
Figure~\ref{fig: 5e8}(b) displays the pressure distribution with the initial design. 
Despite the deformed configuration affecting the wave propagation especially around the region near the traction boundary $\Gamma_t$, almost all of the incident waves reached the outlet. 
The trend in the value of $J$ and the shape change during optimization is similar to the previous case. 
The optimized design was obtained after 119 iterations.
The value of $J$ was $1.03\times 10^{-8}$, which is smaller than the initial design. 
Figure~\ref{fig: 5e8}(c) shows the pressure distribution with the optimized design. 
It is noteworthy that the optimized design successfully achieved low transmission through the outlet under deformation. 

\begin{figure}[htbp]
	\begin{center}
		\includegraphics[scale=0.55]{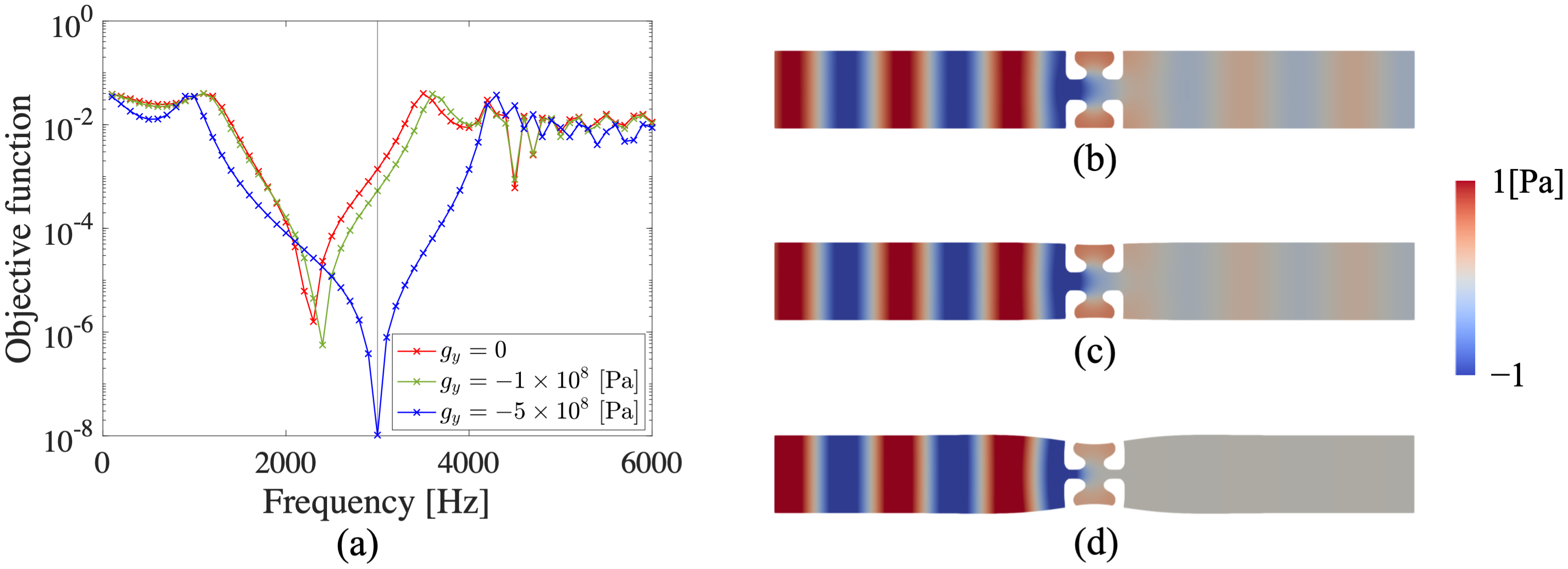}
	\end{center}
	\caption{(a) Frequency responses of $J$ of the optimized design for $g_y=-5\times 10^8~$[Pa]. The red, green, and blue plots represent the frequency responses when $g_y=0$, $g_y=-1\times 10^8~$[Pa], and $g_y=-5\times 10^8~$[Pa], respectively. 
    (b)--(d) Acoustic pressure distribution at 3000 [Hz] with the optimized design for $g_y=-5\times 10^8~$[Pa] when (b) $g_y=0$, (c) $g_y=-1\times 10^8~$[Pa], and (d) $g_y=-5\times 10^8~$[Pa].
    }
	\label{fig: Objfreq_opt5e8}
\end{figure}
Figure~\ref{fig: Objfreq_opt5e8}(a) displays the frequency responses of $J$ for the optimized design with $g_y=-5\times 10^8~$[Pa]. 
The blue line indicates that the value of $J$ reached the minimum value at the target frequency. 
When different traction conditions were applied ($g_y=0$ and $g_y=-1\times 10^8~$[Pa]), the peak frequency at which the value of $J$ took the minimum value was shifted from the target frequency, resulting in an increase in the value of $J$ at that frequency. 
These trends were also observed in the pressure distribution at 3000~[Hz], as shown in Figure~\ref{fig: Objfreq_opt5e8}(b)--(d), and were consistent with those of the optimized design without deformation.



\begin{table}[htbp]
	\begin{center}\caption{ Optimized designs and the values of objective functions with different traction conditions.
            (a) Optimized design for $g_y=0$. (b) Optimized design for $g_y=-1\times 10^{8}\mathrm{[Pa]}$. (c) Optimized design for $g_y=-5\times 10^{8}\mathrm{[Pa]}$.
            }
		\begin{tabular}{|l|l|l|l|} \hline
			&\begin{tabular}{c}
                    ~
                    \begin{minipage}{0mm}
                        \centering
                        \includegraphics[scale=0.6]{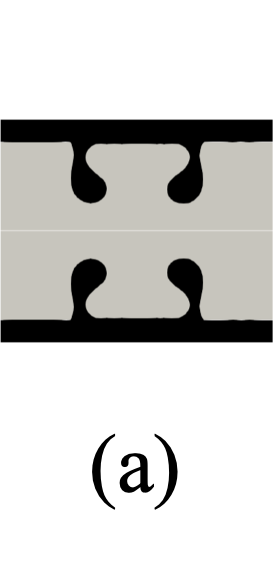}
                    \end{minipage}
			\end{tabular}&
			\begin{tabular}{c}
                    ~
                    \begin{minipage}{0mm}
                        \centering
                        \includegraphics[scale=0.6]{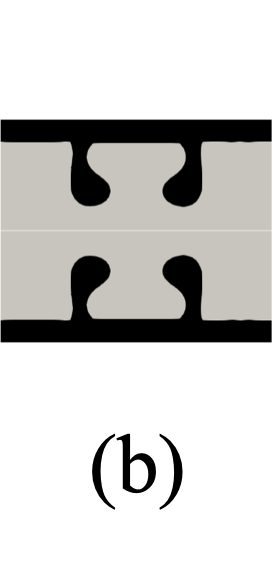}
                    \end{minipage}
			\end{tabular}&
			\begin{tabular}{c}
                    ~
                    \begin{minipage}{0mm}
                        \centering
                        \includegraphics[scale=0.6]{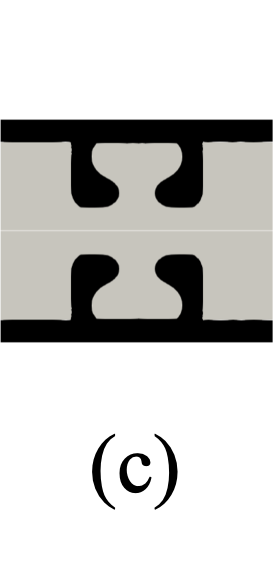}
                    \end{minipage}
			\end{tabular}\\
			\hline 
			$g_y=0$& $9.77\times 10^{-6}$  &$5.66\times 10^{-5}$ &$1.38\times 10^{-3}$\\
			$g_y=-1\times 10^{8}\mathrm{[Pa]}$& $6.76\times 10^{-5}$  &$1.05\times 10^{-6}$ &$5.30\times 10^{-4}$\\
			$g_y=-5\times 10^{8}\mathrm{[Pa]}$& $1.17\times 10^{-4}$  &$3.25\times 10^{-5}$ &$1.03\times 10^{-8}$\\  		
			\hline
		\end{tabular}
		\label{tab: Opt designs and Objs}
	\end{center}
\end{table}
Based on these results, the obtained optimized designs appear to be specialized for their mechanical situation under the traction force $\bm{g}$. 
The results are summarized in Table~\ref{tab: Opt designs and Objs}, which includes a newly optimized design targeting $g_y = -1\times 10^8~$[Pa]. 
The values of $J$ are presented for each optimized design under different values of $g_y$. 
Each design's $J$ value shows the smallest value for the target value of $g_y$ among the three different conditions.
Therefore, the proposed method successfully provides the optimized design for a given deformation condition. 

\subsection{Case2: Deformable acoustic lens for manipulating reflected waves}
\begin{figure}[h!]
	\begin{center}
		\includegraphics[scale=0.25]{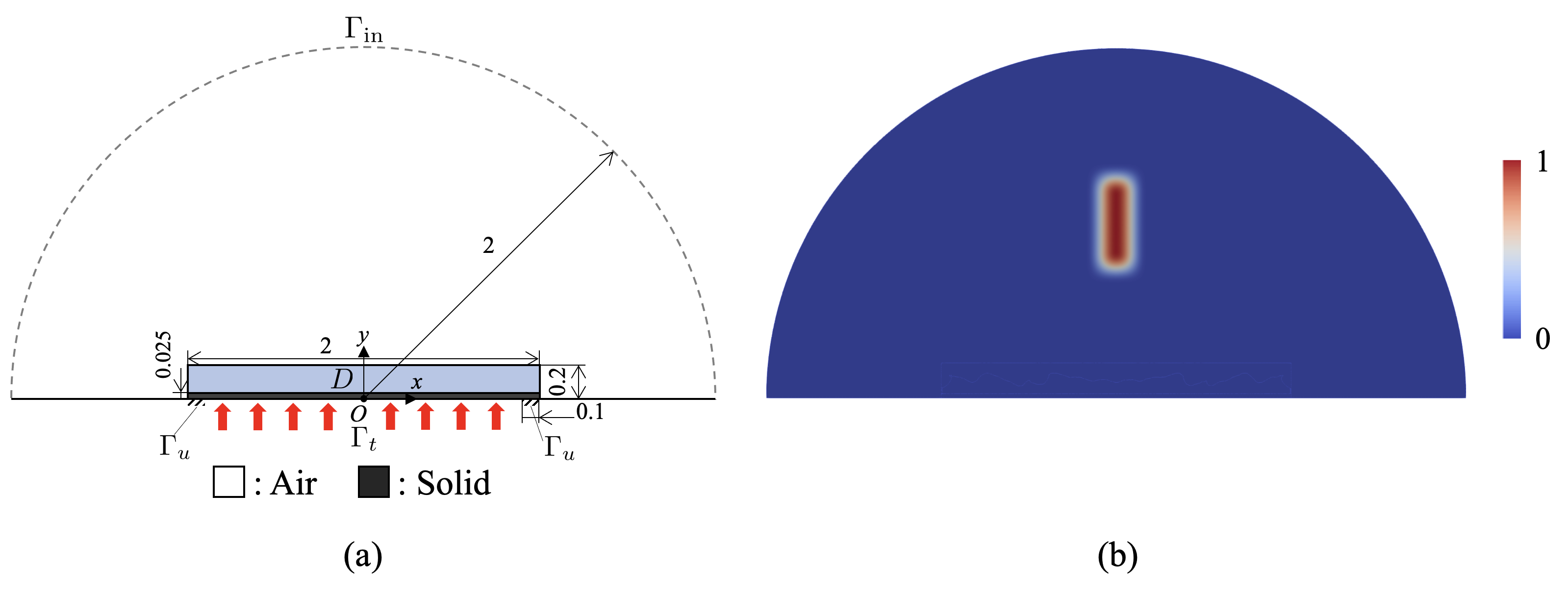}
	\end{center}
	\caption{(a) Setting of design domain. Units are in [m].
                 (b) Distribution of $\tilde{\chi}(\bm{x})$ to define the focal point.}
	\label{fig: DesignSettings}
\end{figure}
In this example, we optimize the shape of acoustic lenses to maximize the amplitude of reflected waves at a desired point while considering the effect of structural deformation. 
The geometry setting of Case 2 is shown in 
Fig.~\ref{fig: DesignSettings}. 
The black-colored region represents a sound-reflective structure placed on a sound-hard wall, which deforms due to the traction force $\bm{g}$ on $\Gamma_t$ with the fixed constraint on $\Gamma_u$. 
Here, we set $\bm{g} = [0, g_y]^T$ with $g_y = 1\times 10^8~$[Pa].
The white region is an air-filled domain, and the boundary $\Gamma_{\mathrm{in}}$ is the input boundary where the incident plane wave is applied with $\bm{k} = [0,-k_b]^T$ and 
a frequency of 2000~[Hz]. 
The blue-colored region in Fig.~\ref{fig: DesignSettings} is the design domain $D$ where the shape of the lens is optimized. 
The parameter $\tau$ to control the perimeter is set to $1\times 10^{-5}$ with the characteristic length $L=0.05$[m]. 
We refine the finite elements such that the size of each element in $D$ is around 0.01 [m]. 
In addition, we set $\epsilon$ in the smoothed Heaviside function $\tilde{H}$ as $\epsilon = 0.01$ to define the profile of $\phi$. 
Furthermore, we set 
$\alpha_e$ in Eq.~(\ref{eq: u_e}) as $\alpha_e=0.05$ and $\alpha_\nu= 0.02$ to extend the velocity field. 
The optimization calculation is halted when the absolute difference between the values of $J$ for two consecutive iterations is smaller than $5\times 10^{-4}$ after 150 iterations of the calculations.

To optimize the structure that realizes acoustic lenses to focus the energy of the reflected waves, we consider minimizing the following two kinds of objective functions:
\begin{align}
    J_1 &= -\int_{\Omega_{\mathrm{all}}}|\pd|^2 \tilde{\chi}(\bm{x}+\uvec) \det \bm{F}(\uvec) d\Omega, \label{eq: obj_Case2}\\
    J_2 &= -\int_{\Omega_{\mathrm{all}}}|p|^2 
    \tilde{\chi}(\bm{x})d\Omega, 
\end{align}
where $\tilde{\chi}(\bm{x})$ is a smooth function that defines the focal point in $\Omega_{\mathrm{all}}$. 
It shows 1 where the amplitude of the acoustic waves is to be maximized and 0 in the other regions. 
$J_1$ measures the performance of the structure with deformation. 
On the other hand, $J_2$ measures the performance without deformation. 
To evaluate the amplitude of the acoustic waves in the spatial coordinate $\bm{x}$, we introduce the displacement field $\uvec$ and the Jacobian $\det \bm{F}(\uvec)$ in the expression of $J_1$. 
Note that $p$ in $J_2$ represents the acoustic pressure without deformation. 

The distribution of $\tilde{\chi}(\bm{x})$ utilized for optimization is presented in Fig.~\ref{fig: DesignSettings}(b). 
This distribution is defined using the smoothed Heaviside function in the following expression:
\begin{align}
    \tilde{\chi}(\bm{x})=
    \tilde{H}(\frac{x-x_0}{R})
    \tilde{H}(-\frac{x-x_1}{R})
    \tilde{H}(\frac{y-y_0}{R})
    \tilde{H}(-\frac{y-y_1}{R})
    .\label{eq: chiplot}
\end{align}
Note that if the smoothing parameter, $\epsilon$, in the definition of $\tilde{H}$ approaches 0, $\tilde{\chi}(\bm{x})$ in Eq.~(\ref{eq: chiplot}) becomes a characteristic function in $\bm{x} \in (x_0,x_1) \times (y_0,y_1)$. 
Here, $R=2~$[m] is utilized to normalize the coordinates in Eq.~(\ref{eq: chiplot}). 
To define $\tilde{\chi}(\bm{x})$ with a smooth distribution, we set $\epsilon=0.05$ in Eq.~(\ref{eq: chiplot}), which is larger than that used for the profile of $\phi$. 
We set $x_0=-\frac{\lambda}{2}$, $x_1=\frac{\lambda}{2}$, $y_0=\frac{R}{2} - \frac{3}{2}\lambda$, and $y_1=\frac{R}{2} + \frac{3}{2}\lambda$, where $\lambda$ is the wavelength in air. 

In the following, we present two numerical examples. 
The first example aims to minimize $J=J_2$, where the design of the acoustic lens is optimized without considering the effect of deformation. 
In the second example, we aim to minimize $J=J_1+J_2$ to optimize the structure with and without structural deformation, enabling for functionalities in both states. 
In this example, we use shape derivatives $DJ_1\cdot \bm{\theta}$ and $DJ_2\cdot \bm{\theta}$ to minimize $J_1$ and $J_2$, respectively. 
These are extended independently by solving Eq.~(\ref{eq: extention_mod}) twice. 
We denote by $v_i$ with $i=1,2$, the extended velocity corresponding to the shape derivative $DJ_i \cdot \bm{\theta}$. 
We then update $\phi$ by using Eq.~(\ref{eq: RDE_weakform}) with a modified normalized velocity, 
$\overline{\nu} = \sum_{i=1}^2 \nu_i (\frac{\int_{D}|\nu_i \delta (u_e)|d\Omega }{\int_{D}|\delta (u_e)|d\Omega})^{-1}$. 

\begin{figure}[h!]
	\begin{center}
		\includegraphics[scale=0.6]{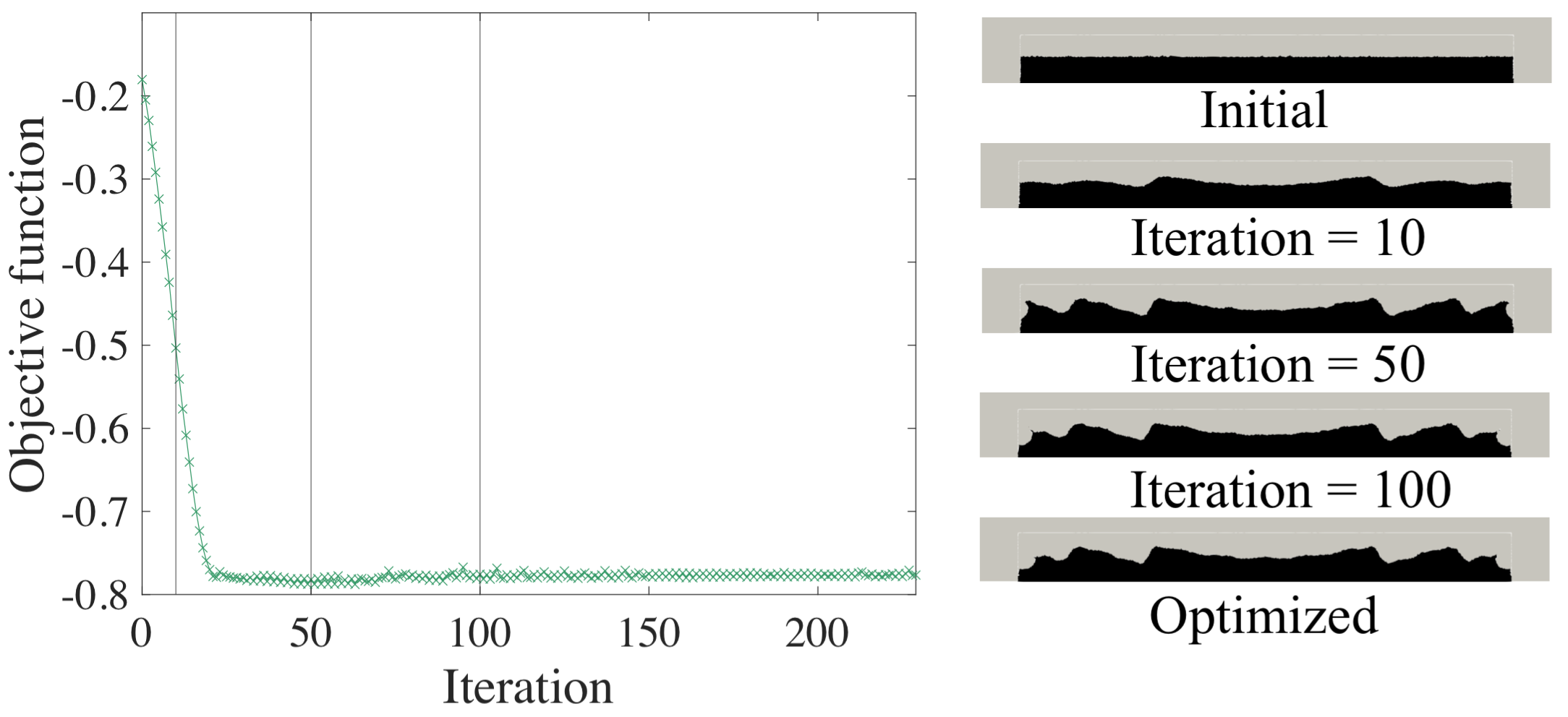}
	\end{center}
	\caption{(Left) History of the objective function $J$ when $g_y=0$. (Right) Initial, intermediate, and optimized designs. }
	\label{fig: obj1_history}
\end{figure}
\begin{figure}[h!]
	\begin{center}
		\includegraphics[scale=0.5]{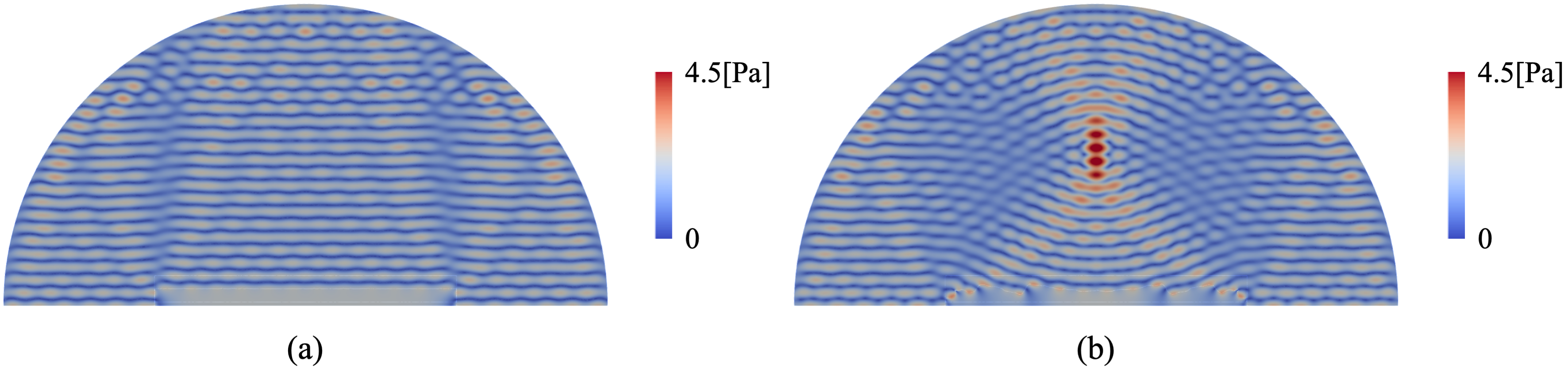}
	\end{center}
	\caption{Distribution of absolute value of acoustic pressure with (a) initial design and (b) optimized design shown in Fig.~\ref{fig: obj1_history}.}
	\label{fig: obj1_p}
\end{figure}

We first present the optimization results for $J=J_2$. 
The history of the objective function and structures during the optimization is shown in Figure~\ref{fig: obj1_history}.
The initial design is a rectangular shape, as illustrated in the right-hand side of Fig.~\ref{fig: obj1_history}.
Figure~\ref{fig: obj1_p}(a) displays the distribution of the absolute value of the acoustic pressure with the initial design. 
The incident plane wave was reflected on the top surface of the initial structure, and the energy of the reflected waves was not concentrated in the target region shown in Fig.~\ref{fig: DesignSettings}(b). 
The value of $J$ with the initial design was estimated as $J=-0.18$. 
As the optimization progressed, the initial planar shape evolved into a shape with multiple steps, similar to that of a Fresnel lens. 
The optimization calculation was stopped at the 229-th iteration, at which the optimized design shown in the  bottom right of Fig.~\ref{fig: obj1_history} was obtained. 
Figure~\ref{fig: obj1_p}(b) shows the distribution of the absolute value of acoustic pressure with the optimized design. 
A large absolute value of the acoustic pressure was observed around the desired region due to the concave surface of the optimized design, which improved the objective function to $J = -0.78$. 

\begin{figure}[h!]
	\begin{center}
		\includegraphics[scale=0.5]{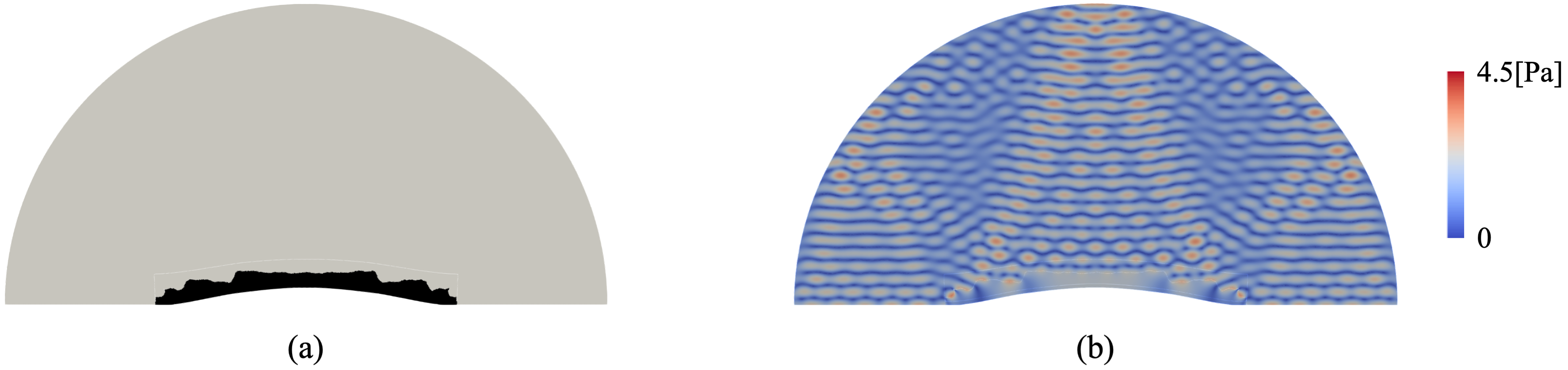}
	\end{center}
	\caption{ 
    (a) Optimized design in Fig.~\ref{fig: obj1_history} deformed by a traction force. (b) Distribution of absolute value of acoustic pressure with the deformed configuration in (a).
     }
\label{fig: obj1_deform}
\end{figure}

However, the performance of the optimized structure deteriorated when a traction force was applied on $\Gamma_t$. 
Figure~\ref{fig: obj1_deform}(a) shows the deformed shape of the optimized design under the traction force, while Fig.~\ref{fig: obj1_deform}(b) illustrates the distribution of the absolute value of acoustic pressure with the deformed configuration. 
Each stepped shape in the optimized design is nearly planar.
Thus, the energy of the reflected waves was not efficiently concentrated in the desired region. 
Consequently, the value of $J$ was $-0.22$ with the deformed structure, which is comparable to the initial value of $J$ without deformation.

\begin{figure}[htbp]
	\begin{center}
		\includegraphics[scale=0.6]{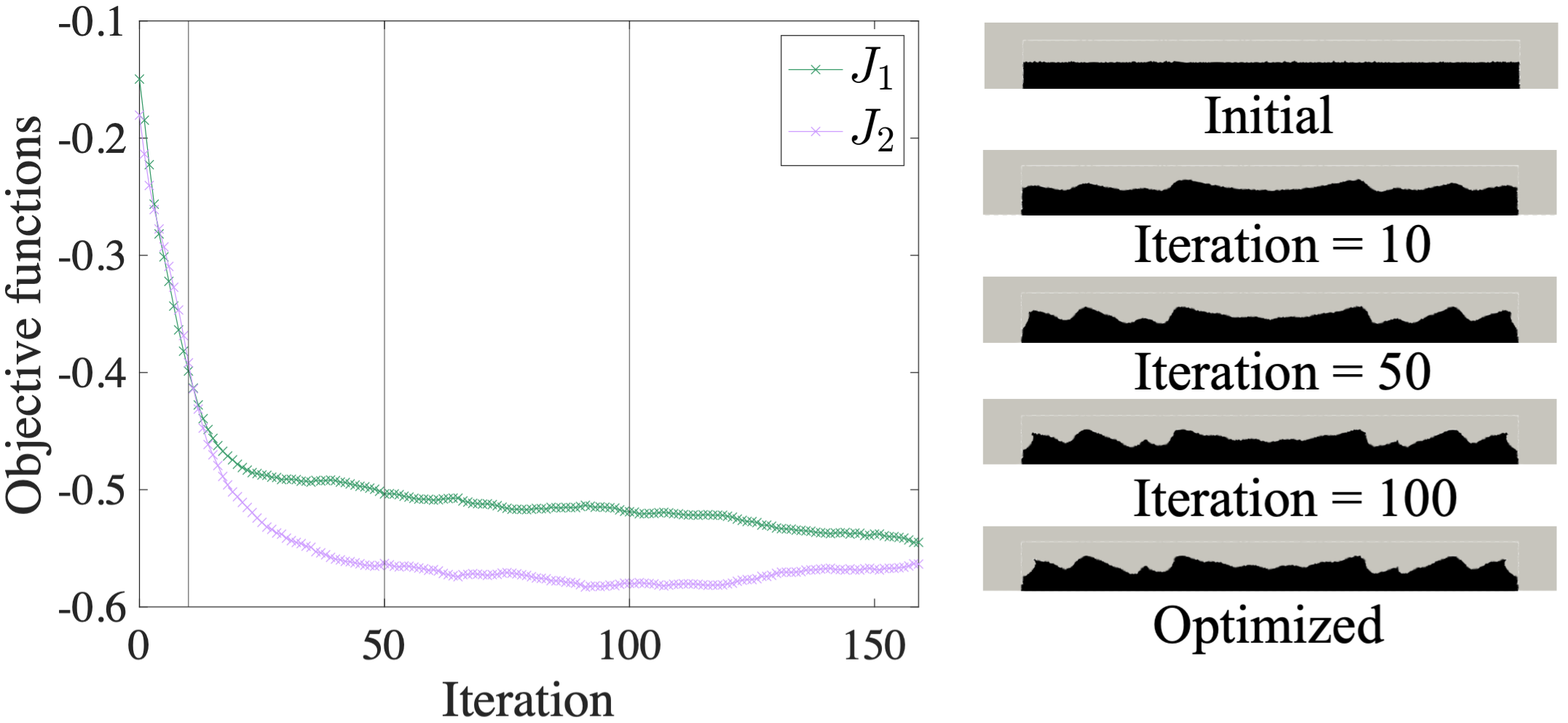}
	\end{center}
	\caption{(Left) History of the objective functions, $J_1$ and $J_2$ when $g_y=1\times 10^8~$[Pa]. (Right) Initial, intermediate, and optimized designs. }
	\label{fig: obj2_history}
\end{figure}

Next, we present the optimization results for $J=J_1 + J_2$. 
Figure~\ref{fig: obj2_history} displays the optimization history of objective functions ($J_1$ and $J_2$) and structures. 
The initial design was identical to the previous example, with estimated objective function values of  $J_1=-0.15$ and $J_2=-0.18$. 
The optimization was terminated at the 159-th iteration, with improved objective function values of $J_1 = -0.54$ and $J_2 = -0.56$. 

\begin{figure}[htbp]
	\begin{center}
		\includegraphics[scale=0.8]{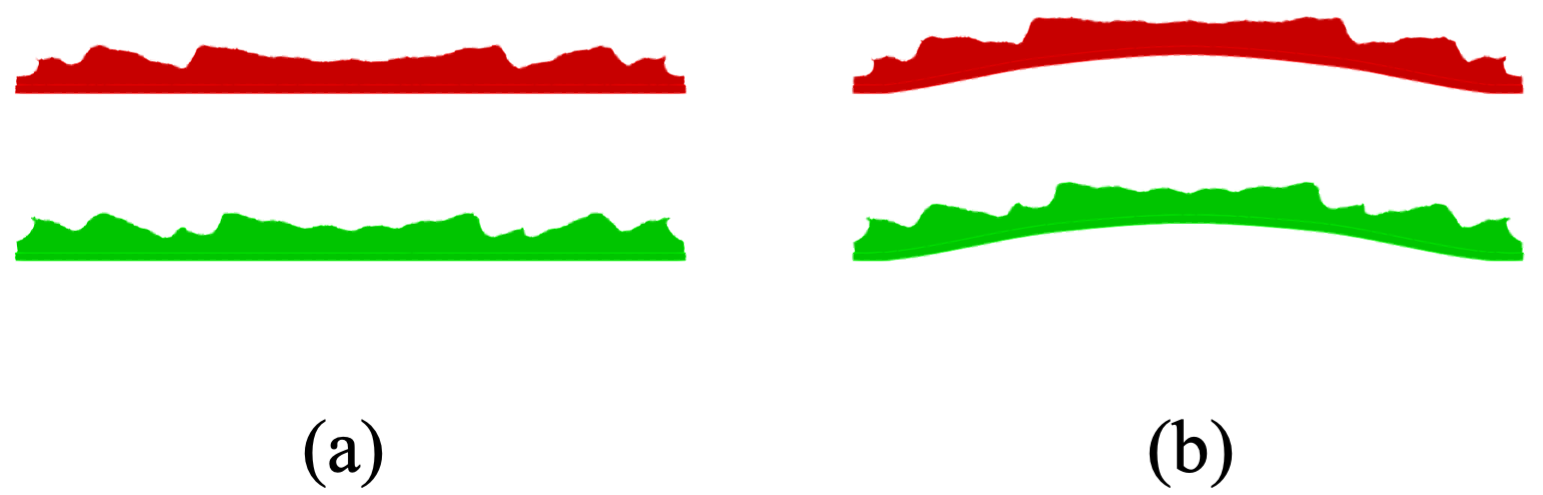}
	\end{center}
	\caption{ Comparison of optimized designs in Fig.~\ref{fig: obj1_history} and Fig.~\ref{fig: obj2_history}.
         (a) Optimized designs when there is no traction force applied to the structure. (b) Deformed optimized designs. 
        The red-colored design represents the optimized shape in Fig.~\ref{fig: obj1_history}, whereas the green-colored one represents that in Fig.~\ref{fig: obj2_history}.
 }
	\label{fig: optconfig_compare}
\end{figure} 

\begin{figure}[h!]
	\begin{center}
		\includegraphics[scale=0.5]{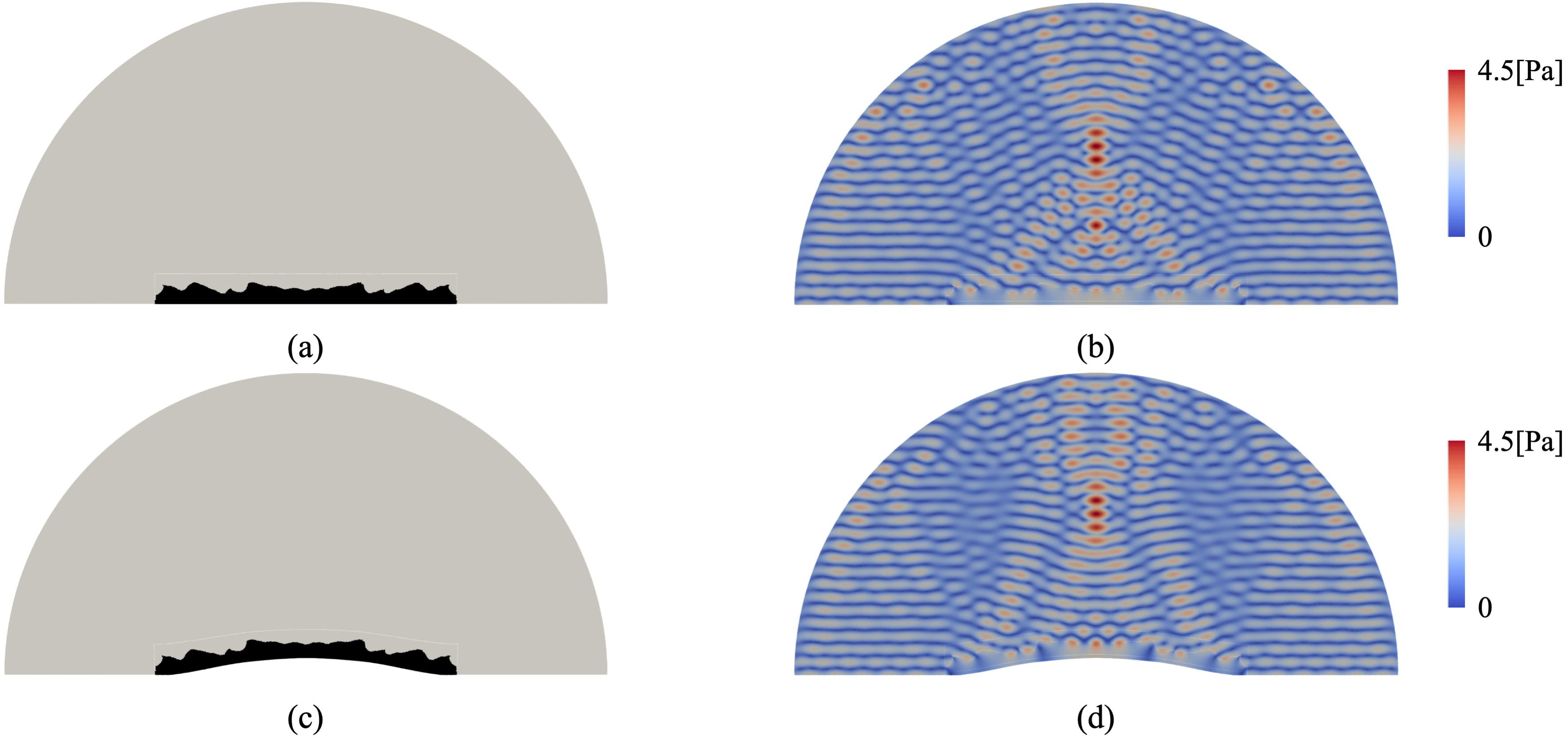}
	\end{center}
	\caption{
 (a) Optimized design in Fig.~\ref{fig: obj2_history} without deformation. (b) Distribution of absolute value of acoustic pressure with the structure shown in (a).
     (c) Optimized design in Fig.~\ref{fig: obj2_history} with the deformation.
     (d) Distribution of absolute value of acoustic pressure with the structure shown in (c).
        }
	\label{fig: obj2_p}
\end{figure}

The optimized structure at the right bottom of Fig.~\ref{fig: obj2_history} has multiple steps, which is similar to the case without deformation. 
However, there is a slight difference in the angle of each step between the shapes in Fig.~\ref{fig: obj1_history} and Fig.~\ref{fig: obj2_history}.
Figure~\ref{fig: optconfig_compare} compares the optimized designs in Fig.~\ref{fig: obj1_history} and Fig.~\ref{fig: obj2_history}. 
Figure~\ref{fig: optconfig_compare}(a) illustrates the two optimized designs when no traction force is applied to the structure. The red-colored design represents the optimized shape in Fig.~\ref{fig: obj1_history}, whereas the green-colored one represents the optimized shape in Fig.~\ref{fig: obj2_history}. 
Figure~\ref{fig: optconfig_compare}(b) shows the deformed optimized designs due to the traction force. 
Although the optimized structure in Fig.~\ref{fig: obj1_history} has almost planar steps, the structure in Fig.~\ref{fig: obj2_history} has a concave surface, which is beneficial for focusing reflected waves when the traction force is applied. 
Figure~\ref{fig: obj2_p} shows the optimized shapes and the distribution of the absolute value of acoustic pressure with and without deformation. 
In the case without deformation, the focal point appears to be beneath the desired region. 
However, the absolute value of pressure is still larger in that region than in other regions due to the multiple steps in the optimized structure. 
In the case with the traction force, the focal point appears to be in the desired region, which improved the value of $J_1$. 
This is because the deformed optimized design also has a concave surface. 
Therefore, we could realize the optimal design of acoustic lenses by considering the structural deformation.

\section{Conclusions}\label{sec: Conclusion}
We proposed a level set-based shape optimization method for designing deformable structures that control acoustic waves. 
The main results are summarized as follows:
	
\begin{enumerate}
    \item We proposed a mathematical model to describe acoustic wave propagation behavior with deformed geometry. This model is based on the Eulerian description with coordinate transformation using the displacement vector field. To perform such a transformation, we introduced a linear elasticity model and extended the displacement field to the air-filled region using the Ersatz approach. 
    \item We formulate the shape optimization problem, with the objective function defined by the acoustic pressure and displacement field. 
    \item We conduct sensitivity analysis to derive the shape derivative for the formulated optimization problem. We followed C\'ea's method to derive the shape derivative and obtained the explicit formula for the shape derivative and the adjoint equations that need to be solved. 
    \item We introduced a level set-based shape and topology optimization method proposed by Feppon et al. \cite{feppon2017introducing} to solve the optimization problem. We also provided some notes on numerical implementation for the optimization, including the optimization flow. 
    \item We provided two kinds of two-dimensional numerical examples to demonstrate the effectiveness of the proposed optimization method. 
    In the first example, we optimized the designs of deformable acoustic tubes exhibiting the best transmission loss under a given traction force.
    In the second example, we optimize the shape of acoustic lenses that can focus the energy of the reflected waves in a target region with and without the traction force. 
    In the second example, we optimized the shape of acoustic lenses that can focus the energy of the reflected waves in a target region with and without the traction force. 
\end{enumerate}
Although our focus was on acoustic waves, our optimization method could be extended to other fields of physics, such as electromagnetic waves. 
In order to describe the deformation of the structure, we introduced the linear elasticity theory for simplicity. However, for larger deformations, nonlinear elasticity theory should be utilized. 
Despite the need for nonlinear elasticity theory in such cases, the proposed optimization framework remains applicable and the shape derivative can still be derived in a similar manner. 
These extensions will be addressed in our future research. 


\appendix
\section{Validation of numerical analyses under a deformation}\label{sec: Validation}

Here, we confirm the validity of the numerical analysis method for acoustic wave propagation problems under structural deformation by comparing the $\pd$ distribution obtained using two different methods. 
The first method uses Eq.~(\ref{eq: Weak form of p' after transformation}) in the FEM analysis, which was also used in the optimization calculation. 
The second method involves using Eq.~(\ref{eq: Weak form of p' before transformation}) in the FEM analysis with the Lagrange mesh. The Lagrange mesh is obtained by moving the nodes in the original mesh with the nodal value of the displacement field $\uvec$. 
We utilized the \texttt{movemesh} command in FreeFEM \cite{MR3043640} to obtain this mesh. 
To evaluate the effectiveness of the proposed method, we compare the pressure distribution at the optimized designs in Case 1 and Case 2 from Section~\ref{sec: numerical examples}. 
To distinguish between the $\pd$ obtained using the proposed method based on Eq.~(\ref{eq: Weak form of p' after transformation}) and the $\pd$ obtained using the Lagrange mesh and Eq.~(\ref{eq: Weak form of p' before transformation}), we denote them as $\pd_{p}$ and $\pd_{l}$, respectively. 

\begin{figure}[h!]
	\begin{center}
		\includegraphics[scale=0.65]{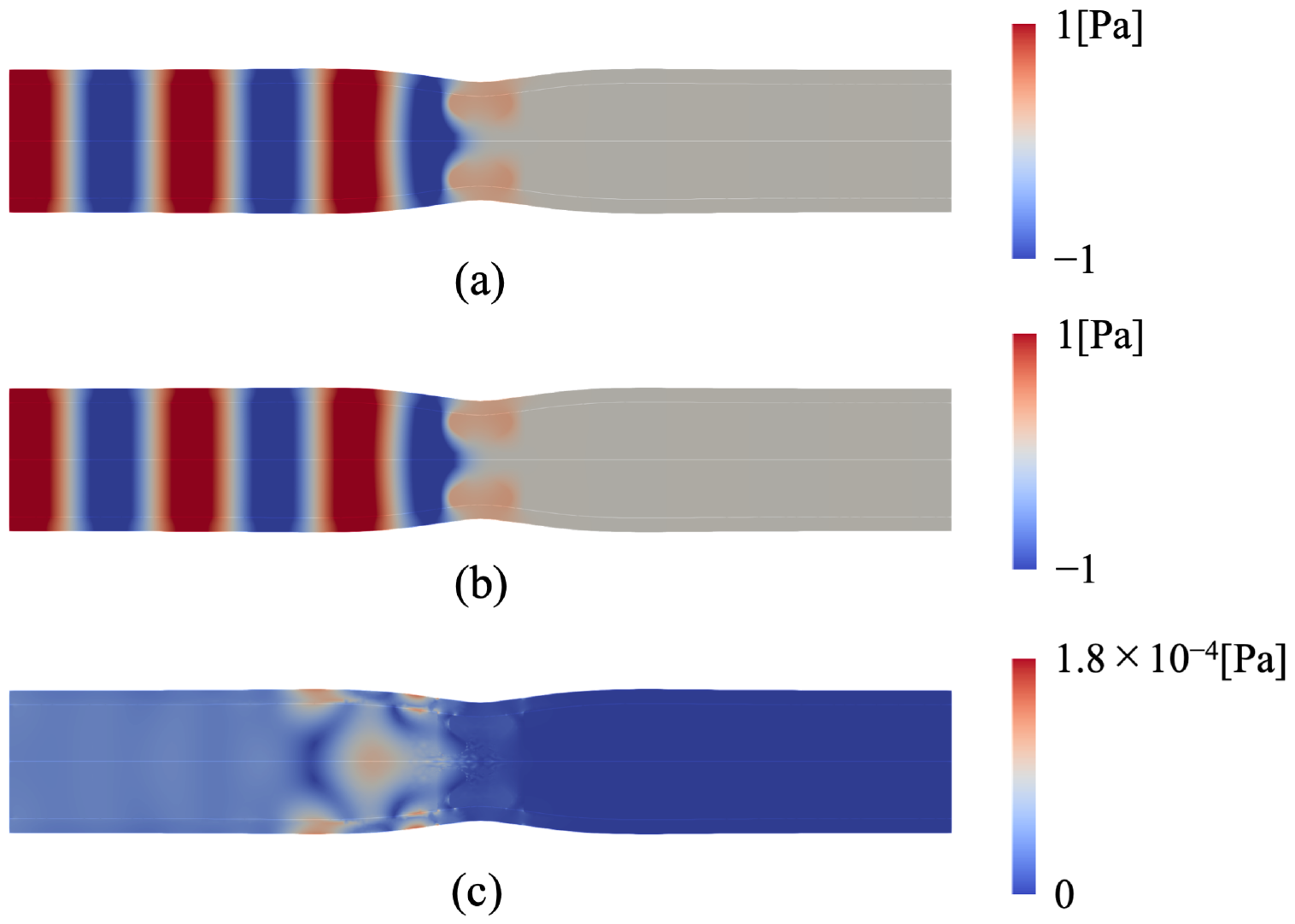}
	\end{center}
	\caption{ Comparison of acoustic pressure distribution for the optimized structure in Fig.~\ref{fig: 5e8} when the traction force with $g_y = -5\times 10^{8}~$[Pa] is applied. 
                 (a) Distribution of $\mathrm{Re}(\pd_p)$.
                 (b) Distribution of $\mathrm{Re}(\pd_l)$. 
                 (c) Distribution of $|\pd_p-\pd_l|$. }
	\label{fig: validity}
\end{figure}

Figure~\ref{fig: validity} depicts a comparison of the acoustic pressure distributions for the optimized design in Fig.~\ref{fig: 5e8}. 
Specifically, Fig.~\ref{fig: validity}(a) displays the distribution of $\mathrm{Re}(\pd_{p})$, while Fig.~\ref{fig: validity}(b) shows that of $\mathrm{Re}(\pd_{l})$ when a traction force with $g_y = -5\times 10^{8}~$[Pa] is applied. 
Although the Lagrange mesh was not used to obtain $\pd_p$, the nodes in (a) were moved using the nodal value of $\uvec$ to compare the result with the distribution of $\mathrm{Re}(\pd_{l})$ shown in (b).  
The pressure distributions in (a) and (b) appear almost identical to each other. 
Figure~\ref{fig: validity}(c) represents the distribution of $|\pd_{p} - \pd_{l}|$, which exhibits much smaller values compared to the amplitude of the incident wave (1~[Pa]). The maximum value of $|\pd_{p} - \pd_{l}|$ was found to be $1.8\times 10^{-4}~$[Pa]. 
As a result, the value of the objective function $J$ in Eq.~(\ref{eq: obj in case 1}) was estimated as $1.032\times 10^{-8}$ 
with $\pd_p$, while it was $1.036\times 10^{-8}$ with  $\pd_l$. 
The relative error between these two values is $0.393$ \%, confirming the validity of the analysis.

\begin{figure}[h!]
	\begin{center}
		\includegraphics[scale=0.5]{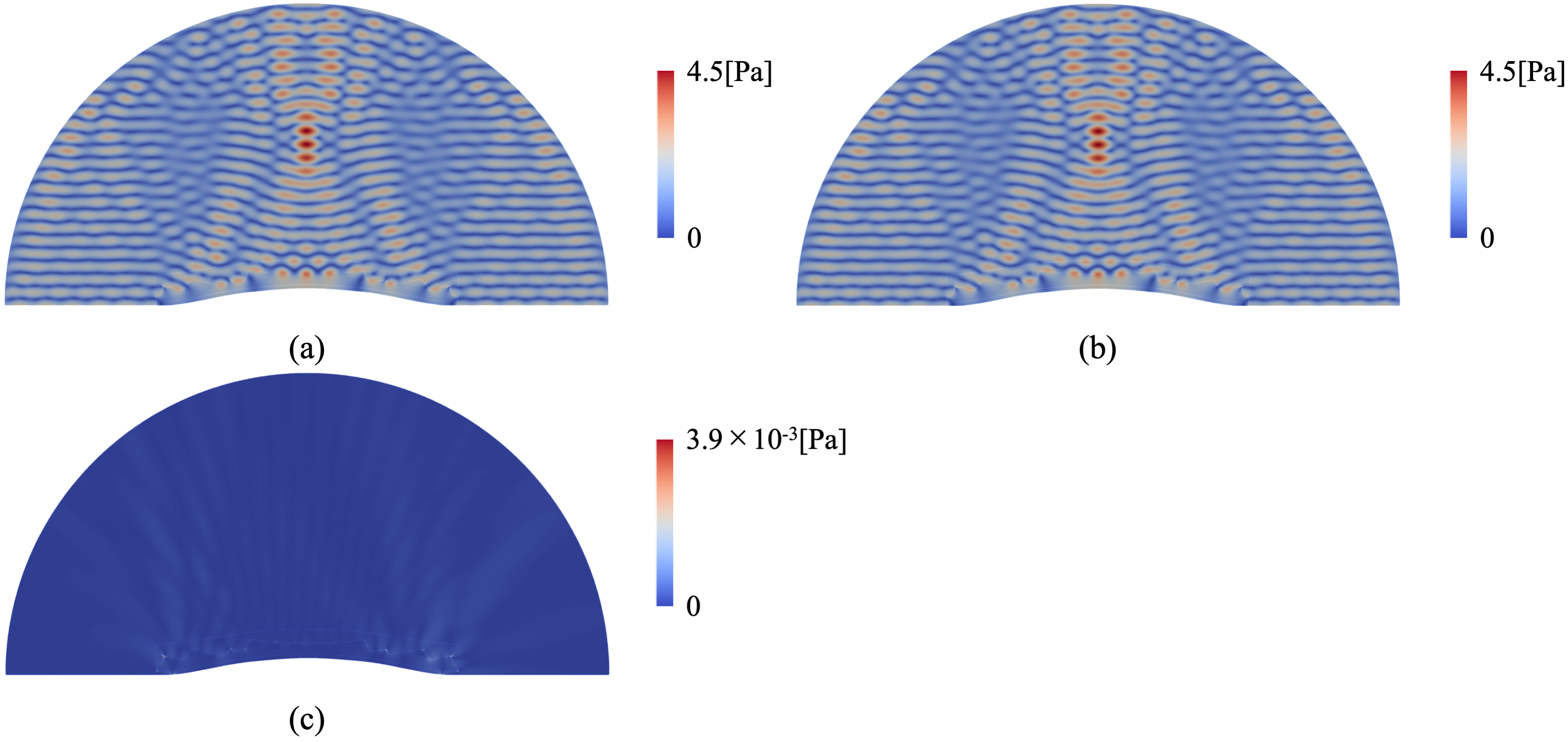}
	\end{center}
	\caption{ Comparison of acoustic pressure distribution for the optimized structure in Fig.~\ref{fig: obj2_history} when the traction force with $g_y = 1\times 10^{8}~$[Pa] is applied. 
                 (a) Distribution of $|\pd_p|$.
                 (b) Distribution of $|\pd_l|$. 
                 (c) Distribution of $|\pd_p-\pd_l|$. }
	\label{fig: obj2_diff}
\end{figure}

Similarly, we compare the pressure distributions in Case 2. 
Figure~\ref{fig: obj2_diff}(a) displays the distribution of $|\pd_p|$, while (b) shows that of $|\pd_{l}|$ when the traction force with $g_y = 1\times 10^{8}~$[Pa] is applied. 
The distribution of the pressure difference $|\pd_p - \pd_l|$ is also shown in Fig.~\ref{fig: obj2_diff}(c). 
Its maximum value, $3.9\times 10^{-3}$, is much smaller than the amplitude of the incident wave (1~[Pa]), which results in almost no difference in the distributions shown in (a) and (b). 
Furthermore, we compare the values of $J_1$ in Eq.~(\ref{eq: obj_Case2}) estimated by $\pd_p$ and $\pd_l$. 
It should be noted that $J_2$ is not a functional of $\pd$, but rather of $p$. 
Therefore, only the values of $J_1$ are the target of this analysis. 
The value of $J_1$ with $\pd_p$ was $-0.544961$, whereas $J_1$ with $\pd_l$ was $-0.544956$. 
Their small relative error ($9.18\times 10^{-4}$\%) demonstrates the validity of the analysis method. 

\bibliographystyle{elsarticle-num} 
\bibliography{reference}


%
%
%
\end{document}